\documentclass[apj,twocolumn,numberedappendix]{emulateapj}

\usepackage{amsmath}
\usepackage{amssymb}
\usepackage{epsfig}
\usepackage{graphicx}
\usepackage{multirow}
\usepackage{longtable}
\usepackage{bbm}
\usepackage{dcolumn}
\usepackage[squaren]{SIunits}
\usepackage{natbib}

\newcommand{\bi}{\begin{itemize}}
\newcommand{\ei}{\end{itemize}}

\newcommand{\be}{\begin{equation}}
\newcommand{\ee}{\end{equation}}
\newcommand{\bea}{\begin{eqnarray}}
\newcommand{\eea}{\end{eqnarray}}

\newcommand{\ie}{{\it i.e.}}

\newcommand{\eg}{{\it e.g.}}

\newcommand{\cf}{{\it cf.}}

\newcommand{\eq}{equation}

\newcommand{\fig}{Figure}

\newcommand{\Sec}{Section}

\newcommand{\App}{Appendix}

\newcommand{\Tab}{Table}

\newcommand{\equ}[1]{\eq~(\ref{equ:#1})}
\newcommand{\figu}[1]{\fig~\ref{fig:#1}}

\begin{document}

\title{UHECR escape mechanisms for protons and neutrons from GRBs,
and the cosmic ray-neutrino connection} 

\author{Philipp Baerwald}
\email{philipp.baerwald@physik.uni-wuerzburg.de}
\author{Mauricio Bustamante}       
\email{mauricio.bustamante@physik.uni-wuerzburg.de}, 
\author{Walter Winter}
\email{winter@physik.uni-wuerzburg.de}

\affil{Institut f{\"u}r Theoretische Physik und Astrophysik, \\ Universit{\"a}t W{\"u}rzburg, 
       97074 W{\"u}rzburg, Germany; \email{winter@physik.uni-wuerzburg.de}}


\begin{abstract}

The paradigm that gamma-ray burst (GRB) fireballs are the sources of the ultra-high energy cosmic rays (UHECRs) is being probed by neutrino observations. Very stringent bounds can be obtained from the cosmic ray (proton)--neutrino connection, assuming that the UHECRs escape as neutrons. In this study, we identify three different regimes as a function of the fireball parameters: the standard ``one neutrino per cosmic ray'' case, the optically thick (to neutron escape) case, and the case where leakage of protons from the boundaries of the shells (direct escape) dominates. In the optically thick regime, photomeson production is very efficient, and more neutrinos will be emitted per cosmic ray than in the standard case, whereas in the direct escape-dominated regime, more cosmic rays than neutrinos will be emitted. We demonstrate that, for efficient proton acceleration, which is required to describe the observed UHECR spectrum, the standard case only applies to a very narrow region of the fireball parameter space. We illustrate with 
several observed examples that conclusions on the cosmic ray--neutrino connection will depend on the actual burst parameters. We show that the definition of the pion production efficiency currently used by the IceCube collaboration underestimates the neutrino production in the optically thick case. Finally, we point out that the direct escape component leads to a spectral break in the cosmic ray spectrum emitted from a single source. The resulting ``two-component model'' can be used to even more strongly pronounce the spectral features of the observed UHECR spectrum than the dip model.

\end{abstract}


\section{Introduction}

Cosmic ray observations tell us that particles with energies higher than $10^{19} \, \electronvolt$ hit the Earth, which are expected to be of extragalactic origin. The search for the sources of these ultra-high energy cosmic rays (UHECRs) is therefore one of the main objectives in high-energy astrophysics. It can be either performed directly, by cosmic ray observations, or indirectly, by looking for the neutrinos accompanying the cosmic ray emission. So far, no evidence for a correlation between specific UHECR sources and cosmic ray measurements has been found. 

One class of potential UHECR sources are gamma-ray burst (GRB) fireballs (see \citet{Piran:2004ba} or \citet{Meszaros:2006rc} for reviews), where the cosmic rays are expected to be accelerated to the highest energies by collisions with the interstellar medium \citep{Rees:1992ek}, or by internal collisions inside the ejected material \citep{Paczynski:1994uv,Rees:1994nw}. The general fireball model describes a GRB as a catastrophic release of energy, during which matter of the order of a solar rest mass is ejected from a compact object. These ejecta are then first accelerated to ultra-relativistic speeds, and are then assumed to coast at constant velocity while expanding into interstellar space. It is assumed that the bursts can also lose energy via radiation at higher radii during the expansion of the fireball. The expansion ends when the ejecta hit the interstellar medium and are decelerated. The observations in several energy bands have shown that GRBs have several distinct phases of emission, with the prompt 
emission phase being the most energetic one. During this phase, the burst is mainly visible in gamma-rays, and the emission is assumed to originate from the coasting phase of the fireball. The high variability and non-thermal properties of the observed gamma-ray spectra give rise to the notion that the prompt emission might be due to the collision of internal shells, leading to Fermi shock acceleration of the charged particles such as electrons or protons (internal shock model).
While recent observations point towards a heavier composition at the highest energies~\citep{Abraham:2010yv}, we focus on protons as candidates for the UHECRs in this study, for which plausible models for the particle acceleration and emission from GRBs exist.
Especially the idea of indirect escape of protons, via neutrons, from the shells has been a very popular extension of the internal shock model in the literature. However, there are also several other alternative models for GRB emission which describe certain aspects of the observed emission well, such as magnetic reconnection models \citep{Lyutikov:2003ih,Zhang:2010jt,Zhang:2012qy} or photospheric emission models \citep{Rees:2004gt,Giannios:2007yj,Murase:2008sp,Wang:2008zm,Beloborodov:2009be,Gao:2012ay,Lazzati:2013ym}.

Because of the high photon densities, it is expected that the protons accelerated in the colliding shocks dissipate energy into pion production. 
In the standard picture, this can be described  by the $\Delta(1232)$-resonance 
\begin{equation}
	p + \gamma \rightarrow \Delta^+ \rightarrow \left\{\begin{array}{lc} n + \pi^+ & \frac{1}{3} \text{ of all cases} \\[0.2cm]  p + \pi^0 & \frac{2}{3} \text{ of all cases} \end{array} \right.  . \label{equ:Delta}
\end{equation}
A substantial neutrino flux then originates from $\pi^+$ decays via the decay chain
\begin{eqnarray}
\pi^+ & \rightarrow & \mu^+ + \nu_\mu \, ,\nonumber \\
& & \mu^+ \rightarrow e^+ + \nu_e + \bar{\nu}_\mu \, , \label{equ:piplusdec}  
\end{eqnarray}
where $\nu_e:\nu_\mu:\nu_\tau$ are produced in the ratio $1:2:0$. 
On the other hand, the neutrons decay via
\begin{equation}
n \rightarrow  p + e^- + \bar{\nu}_e  \label{equ:ndec}  \, ,
\end{equation}
typically outside the source, which leads to a cosmic ray (proton) flux even if the protons themselves are magnetically confined (``neutron model''), as discussed in, \eg, \citet{Mannheim:1998wp}.
Highly energetic gamma-rays originating from the $\pi^0$ decays are injected into the electromagnetic cascade, which leads to constraints from the {\it Fermi}-LAT diffuse GRB measurement; see, \eg, \citet{Ahlers:2010fw} or \citet{Li:2012gf}. 

The neutrino flux has been predicted for the standard internal shock model in \citet{Waxman:1997ti}, assuming that GRBs are the sources of the UHECRs. If one assumes that the observed gamma-ray spectrum represents the photon density within the source in the prompt phase, one can calculate the expected neutrino fluence from the observed gamma-ray fluence; corresponding analytical methods have been developed in \citet{Guetta:2003wi}, \citet{Becker:2005ej}, and \citet{Abbasi:2009ig}. Recently, the IceCube collaboration has however strongly constrained the neutrino flux from GRBs, see \citet{Abbasi:2011qc,Abbasi:2012zw}, with the conclusion that these simple approaches are already severely constrained. Nonetheless, it is known that additional photomeson production processes somewhat harden the neutrino spectra, and that the cooling of the secondaries and flavor mixing change the spectral shape and flavor composition, see \citet{Kashti:2005qa}, \citet{Murase:2005hy}, \citet{Lipari:2007su}, \citet{Hummer:2010vx}, and \citet{Baerwald:2010fk}. In addition, the predicted normalization is significantly reduced if spectral 
effects on the pion production efficiency, the energy dependence of the mean free path of the protons, and the impact of the secondary cooling on the energy budget are taken into account, see \citet{Hummer:2011ms}, \citet{Li:2011ah}, and \citet{He:2012tq}. If one extrapolates a quasi-diffuse flux from a few GRBs, the low statistics of the stacking sample will lead to a systematical error~\citep{Baerwald:2011ee}. Given the astrophysical and systematical uncertainties of the model in \citet{Waxman:1997ti}, the current neutrino observations just start to enter the predicted neutrino flux range, and the full-scale IceCube experiment should find neutrinos after ten years of operation if the baryonic loading of the jets is as high as anticipated in \citet{Hummer:2011ms}. Note, however, that if the prompt emission comes from larger radii, no neutrinos may be found, see \citet{He:2012tq}. Additionally, there have been recent efforts to calculate the neutrino emission from GRBs in the dissipative photospheric models~\citep{Gao:2012ay}, as well as efforts for model-independent calculations~\citep{Zhang:2012qy}.

As far as the direct connection between neutrinos and cosmic rays is concerned, \equ{Delta} suggests roughly one muon neutrino per cosmic ray after flavor mixing, which changes the $\nu_e:\nu_\mu:\nu_\tau$ from $1:2:0$ to $1:1:1$. This hypothesis has been tested in \citet{Ahlers:2011jj} and \citet{Abbasi:2012zw}, with the conclusion that GRBs cannot be the sole source of the UHECR protons. In a more general framework, the authors of \citet{Kistler:2013my} conclude that the protons resulting from  photopion processes are not sufficient to explain the cosmic-ray measurements. Therefore, we discuss the validity of the assumptions going into the ``one neutrino per cosmic ray'' paradigm, henceforth called ``the standard case'':
\begin{enumerate}
\item
 The protons are magnetically confined, and cosmic rays can only escape as neutrons.
\item
 The protons interact only once, at most, and the produced neutrons can escape from the source (source optically thin to neutron escape).
\end{enumerate}
If one of these assumptions is violated, the consequences are obvious: protons ``leaking'' are not accompanied by neutrino production. On the other hand, multiple interactions will enhance the neutrino flux compared to the standard picture, while only neutrons from the boundaries can escape. In this study, we will explore these two regimes in addition to the standard picture, and we will demonstrate that, for high proton acceleration efficiencies, which are required to describe the observed UHECR spectrum, the standard case only occupies a very small region of the parameter space.

This paper is organized as follows: We describe the implementation of the GRB fireball model for our purpose in \Sec~\ref{sec:model}.  This section may be skipped if the reader is familiar with this or similar models. In \Sec~\ref{sec:direct}, we discuss the direct escape of cosmic rays from optically thin (to neutron escape) sources, and we comment on the additional effects of diffusion in \App~\ref{app:diffusion}. On the other hand, we treat the optically thick (to neutron escape) regime in \Sec~\ref{sec:thick}, where we also comment on the pion production efficiency in that regime in \App~\ref{app:pion}. In \Sec~\ref{sec:space} we relate the important cases for the cosmic ray-neutrino connection to regions in parameter space, and we discuss specific (observed) examples in \Sec~\ref{sec:spec}. Furthermore, we illustrate in
\Sec~\ref{sec:cr} the impact of an additional escape component on the observed cosmic ray flux. Finally, we summarize in \Sec~\ref{sec:summary}. 

\section{Implementation of the GRB fireball model}
\label{sec:model}

We use a simplified description of the relativistically expanding fireball, based on \citet{Waxman:2003vh}, to illustrate our main points. Primed quantities refer to the shock rest frame (SRF), and unprimed quantities to the observer's or source (cosmologically co-moving) frame, which we clearly indicate. 
GRB observations exhibit a strong time variability over a scale $t_v$ (defined in the observer's frame), which can be related to a basic length scale $r_0 = c \, t_v/(1+z)$ in the source frame.  We assume that the central engine of the GRB emits shells of thickness $\Delta r \simeq c \cdot t_v/(1+z) = r_0$ in the source frame, since causality implies that variations of the timescale $t_v$ can only propagate over a distance scale $\Delta r$. The time evolution of the fireball can be divided into different zones. In the first zone,  the shell gets accelerated, powered by the energy transfer from the thermal photons to the baryons in the shell. The Lorentz factor of the shell grows with the radius until a maximum value $\Gamma$ is reached, which is, in principle, given by $\Gamma = E_{\text{tot}} / (Mc^2)$, where $M$ is the total mass of baryons and $E_\text{tot}$, the total energy of the fireball. This transition is complete at a 
radius $r \approx \Gamma \, r_0$. Here the second zone is considered to start: since the shell is accelerated to its maximal velocity, it coasts with constant $\Gamma$, while the expansion of the width of the shell itself is still negligible. However, when the shell reaches the radius $r \approx \Gamma^2 \, r_0$, the growth of the shell width can no longer be neglected, since $\delta \Delta r \simeq r/\Gamma^2 \simeq r_0$. We will come back to this later when we discuss the effects of an expanding shell. For now, we assume that the shell width in the SRF is roughly given by
\begin{equation}
 \Delta r' \simeq \Gamma \, c \, \frac{t_v}{1+z}
\label{equ:dr}
\end{equation}
at the indicated radius.
At roughly the same radii,
\begin{equation}
r_C \simeq 2 \, \Gamma^2 \, r_0  = 2 \, \Gamma^2 \, c \, \frac{t_v}{1+z} , \label{equ:rc}
\end{equation}
the collisions of the different shells start, based on the assumed fluctuations of the shells Lorentz factors of the order of $\Delta \Gamma / \Gamma \sim 1$. 
External collisions with the interstellar medium can also lead to efficient proton acceleration, which we do not consider since the typical photon densities are orders of magnitude lower than in the internal collision zone.

We focus on the description of the prompt phase, which is associated with the collisions of the shells. 
Since a relativistically expanding fireball may undergo different phases in its expansion, with varying parameters, we describe the physics of one collision, following \citet{Baerwald:2011ee}, \citet{Hummer:2011ms}, and \citet{Winter:2012xq} and consistent with \citet{Waxman:1997ti} and \citet{Guetta:2003wi}. If one assumes that the collisions occur at the same radius $r_C$, as it is implied in all of the state-of-the-art neutrino analyses~\citep{Abbasi:2011qc,Abbasi:2012zw}, the total fluences can be obtained by summing over 
$N \simeq T_{90}/t_v$ such collisions, where $T_{90}$ is the time during which 90\% of the total energy is observed. Our shell-dependent approach has the advantage that the conventional results can be easily retrieved, and that in addition the relation to collision radius-dependent models can be established. 

For the photohadronic interactions, the secondary (such as pion) injection $Q'(E')$ (in units of $\giga\reciprocal\electronvolt \, \centi\meter\rpcubed \, \reciprocal\second$) can be computed from the proper  photon $N'_{\gamma}(\varepsilon')$ and proton $N'_p(E'_p)$ densities (SRF, in units of $\giga\reciprocal\electronvolt \, \centi\meter\rpcubed$) as 
\begin{equation}
Q'(E') = \int\limits_{E'}^{\infty} \frac{dE_p'}{E_p'} \, N_p'(E_p') \, \int\limits_{0}^{\infty} c \, d\varepsilon' \, N_\gamma'(\varepsilon') \,  R( x,y )  \, .
\label{equ:prodmaster}
\end{equation}
Here, $x \equiv E'/E_p'$ is the fraction of energy going into the secondary particles, $y \equiv (E_p'\varepsilon')/m_pc^2$, and $R( x,y )$ is a ``response function''. If many interaction types are considered, this response function can be quite complicated. Nevertheless, if it is known from particle physics,  \equ{prodmaster} can be used to compute the secondary injection for arbitrary proton and photon spectra; see \citet{Hummer:2010vx}.
Note that the secondary injection depends on the product normalization of the proton density $N_p'(E_p')$ and the target photon density $N_\gamma'(\varepsilon')$. 
Once the proper proton and photon densities (including the spectral shapes) are known, as well as the magnetic field $B'$, the secondary meson and neutron production is just a straightforward particle physics consequence. We use the method from \citet{Hummer:2010vx}, based on the physics of SOPHIA~\citep{Mucke:1999yb}, for the computation of the photohadronic interactions. For the secondary meson decays (including the helicity dependence of the muon decays), see, \eg, \citet{Lipari:2007su}. The magnetic field effects and flavor mixing are included as in \citet{Baerwald:2011ee} and \citet{Hummer:2011ms}. Below, we will describe how to determine the relevant input $N_p'(E_p')$, $N_\gamma'(\varepsilon')$, and $B'$ from the observables.

In \equ{prodmaster}, two types of spectra are present: the injection/ejection spectrum $Q'$ and the steady spectrum $N'$. For a specific particle species, these are related to each other by a kinetic equation describing energy losses and escape. If the energy losses can be neglected, they are, for one species of particles, related by
\begin{equation}
 \label{equ:steadystateesc}
 N'(E')=Q'(E')\,t'_\mathrm{esc} \, ,
\end{equation}
where $t'_\mathrm{esc}$ is the escape time. For example, the observed gamma-ray spectrum can be obtained from $Q_\gamma'$, whereas the spectrum relevant for the photohadronic interactions in \equ{prodmaster} is $N_\gamma'$. Typically, one establishes a relationship between observed gamma-ray fluence and target photon density by implying that the gamma-rays escape over $t'_\mathrm{esc}=t'_\mathrm{dyn} \simeq \Delta r'/c$, which means that \equ{steadystateesc} can be used. However, if the optical thickness to pair production or other processes is of order unity, this assumption does not apply, and the observed spectrum is not representative for the density in the source anymore.

For $N_\gamma'(\varepsilon')$, a broken power law is normally assumed, parameterized as
\begin{equation}
	N'_{\gamma}(\varepsilon') \propto \left\{ \begin{array}{ll} \left( \frac{\varepsilon'}{\varepsilon'_{\gamma,\text{break}}} \right)^{-\alpha_{\gamma}} & \varepsilon'_{\gamma,\text{min}} \leq \varepsilon' < \varepsilon'_{\gamma,\text{break}} \\ \left( \frac{\varepsilon'}{\varepsilon'_{\gamma,\text{break}}} \right)^{-\beta_{\gamma}} & \varepsilon'_{\gamma,\text{break}} \leq \varepsilon' < \varepsilon'_{\gamma,\text{max}} \\ 0 & \text{else} \end{array} \right.  \label{equ:targetphoton}
\end{equation}
with  $\varepsilon'_{\gamma,\text{break}} = \mathcal{O}(\kilo\electronvolt)$ the break energy of the photon spectrum in the SRF. Typical values for the spectral indices are $\alpha_{\gamma} \approx 1$ and $\beta_{\gamma} \approx 2$. The minimal and the maximal photon energies are chosen to be $\varepsilon'_{\gamma,\text{min}} = 0.2 \, \electronvolt$ and $\varepsilon'_{\gamma,\text{max}} = 300 \cdot \varepsilon'_{\gamma,\text{break}}$ in our calculations, if the highest energetic photons can escape (see below). These values are far enough away from the break energy to have no visible effect on the predicted neutrino spectra~\citep{Lipari:2007su,Baerwald:2011ee}, though they can somewhat affect the neutron escape spectra; see \Sec~\ref{sec:spec} for a more detailed discussion.
In addition, the energy partition is hardly affected by the maximal photon energy for a spectral index $\beta_\gamma \gtrsim 2$, since the energy in photons then depends only logarithmically on the maximal photon energy, at most.
Note that high-energy photons will not be able to escape above the pair production threshold. In this case, we choose $\varepsilon'_{\gamma,\text{max}}$ consistent with the pair production threshold.\footnote{
We use \eq~(6) from \citet{Waxman:1997ti} to estimate that, applicable for the $\varepsilon^{-2}$-spectra (above the break), assuming that the photon spectrum extends to infinitely high energies.  This is only a rough estimate, since gamma-rays may interact by additional processes. The impact on $\varepsilon'_{\gamma,\text{max}}$ is, however, typically small. }

In case of the internal collisions, it is generally assumed that Fermi shock acceleration leads to a non-thermal particle spectrum of the form
\begin{equation}
	N_p'(E_p') \propto (E_p')^{-\alpha_p} \cdot \exp\left( -\left(\frac{E_p'}{E'_{p,\text{max}}}\right)^k \right) \label{equ:protons}
\end{equation}
with the spectral index $\alpha_p \approx 2$. For the exponential cutoff, we choose $k \simeq 2$ unless noted otherwise. The maximal proton energy $E'_{p,\text{max}}$ can be obtained by comparing the acceleration timescale to the dominant loss timescale
\begin{equation}
t'_{\text{acc}}(E'_{p,\text{max}}) = \text{min} \left[ t'_{\text{dyn}},
t'_{\text{syn}}(E'_{p,\text{max}}),
t'_{p\gamma}(E'_{p,\text{max}}) \right] \, .
\end{equation}
Here we assume that the acceleration time (in Gaussian cgs units) is given by 
\begin{equation}
	t'_{\text{acc}}(E') = \frac{E'}{\eta \, c \, e \, B'} \label{equ:tacc} \, ,
\end{equation}
with the elementary charge $e \simeq 4.803 \cdot 10^{-10} \, \mathrm{Fr}$ and $\eta$ the acceleration efficiency ($\eta$ is defined here so that large values mean efficient acceleration). It is generally assumed that the dominant loss mechanism is either the adiabatic loss due to the expansion of the shell or the synchrotron loss of the protons due to the magnetic fields present in the shells. We do not consider the adiabatic loss timescale explicitly, since we assume that it is of the same order as the dynamical timescale $t'_{\text{dyn}}(E')$.
The synchrotron loss time is given by
\begin{equation}
	t'_{\text{syn}}(E') = \frac{9 \, m^4}{4 \, c \, e^4 \, B'^2 \, E'} \label{equ:tsyn} \, ,
\end{equation}
with the particle mass $m$ being in $\mathrm{erg}$, using the relation $1 \, \mathrm{erg} = 624.15 \, \giga\electronvolt$. Moreover, the photohadronic timescale $t_{p \gamma}'$ is numerically computed from the interaction rate as given in \citet{Hummer:2010vx}.

Let us now derive $N_\gamma'(\varepsilon')$ and $N_p'(E_p')$ from the observables. Frequently used observables are the  (bolometrically corrected) gamma-ray fluence of a detected GRB, $S_{\text{bol}}$ (in units of $\mathrm{erg} \, \centi\meter\rpsquared$), or the radiative flux $F_\gamma$ (in units of $\mathrm{GeV} \, \centi\meter\rpsquared \, \mathrm{s}^{-1}$). Here we focus on a momentary picture of the fireball, described by (a possibly bolometrically corrected) $F_\gamma$, which leads to the isotropic equivalent energy per shell (or collision)
\begin{equation}
 	E^{\text{sh}}_{\text{iso}} \simeq \frac{4\pi \, d_L^2}{(1+\textit{z})} \, F_\gamma  \, t_v \quad , 
 \label{equ:eisobol}
\end{equation} 
where $d_L$ is the luminosity distance. One has  $E'^{\text{sh}}_{\text{iso}}= E^{\text{sh}}_{\text{iso}}/\Gamma$ in the SRF, and $L_{\gamma,\text{iso}} = E^{\text{sh}}_{\text{iso}} \cdot (1+z)/t_v$.
Assuming energy equipartition between photons and electrons,  the photons carry a fraction $\epsilon_e$  (fraction of energy in electrons) of the total energy $E^{\text{sh}}_{\text{iso},\text{tot}}$, and
 \begin{equation}
 	E^{\text{sh}}_{\text{iso},\text{tot}} = \epsilon_e^{-1} \cdot E^{\text{sh}}_{\text{iso}} \, .
 \label{equ:eiso}
 \end{equation}
 In order to compute the photon and proton densities in the SRF, it turns out to be useful to define  an ``isotropic volume'' $V'_{\mathrm{iso}}=4 \pi \, r_C^2 \,  \Delta r' \propto \Gamma^5$, where the latter relationship can be easily read off from Eqs.~(\ref{equ:dr}) and~(\ref{equ:rc}). Here $V'_{\mathrm{iso}}$  can be interpreted as the volume of the interaction region assuming isotropic emission by the source.\footnote{Since both the energy and the volume of the source need to be, in principle, corrected by a beaming factor, this beaming factor cancels in the computation of the energy densities. }
If the characteristics of all collisions are alike, $S_{\text{bol}} \simeq F_\gamma \, T_{90}$.

Now one can determine the normalization of the photon density in \equ{targetphoton} and the proton density  in \equ{protons} from
\begin{equation}
 \int  \, \varepsilon' \, N'_{\gamma}(\varepsilon') \mathrm{d}\varepsilon' = \frac{E'^{\text{sh}}_{\text{iso}}}{V'_{\text{iso}}} \, , \qquad
\int \, E'_p \, N'_p(E'_p) \, \mathrm{d}E'_p =  \frac{1}{f_e} \frac{E'^{\text{sh}}_{\text{iso}}}{ V'_{\text{iso}}}  \, .
 \label{equ:norm}
\end{equation}
Here $f_e$ is the ratio between energy in electrons and protons ($f_e^{-1}$ is the baryonic loading).
Assuming that the magnetic field carries a fraction $\epsilon_B$ of $E'^{\text{sh}}_{\text{iso}}$, one has in addition
\begin{equation}
 U'_B = \frac{\epsilon_B}{\epsilon_e} \cdot \frac{E'^{\text{sh}}_{\text{iso}}}{V'_{\text{iso}}} \quad \text{or} \quad B' = \sqrt{8\pi \, \frac{\epsilon_B}{\epsilon_e} \cdot \frac{E'^{\text{sh}}_{\text{iso}}}{ V'_{\text{iso}}}}  \, .
\label{equ:B}
\end{equation}
After photohadronic interactions and weak decays, one obtains the injection spectrum of secondary neutrinos or neutrons $Q'$, which is to be translated into the observable neutrino or neutron fluence $\mathcal{F}^{\text{sh}}$ (in units of $\mathrm{GeV^{-1} \, cm^{-2}}$) per shell:
\begin{equation}
 \mathcal{F}^{\text{sh}} = t_v \, V'_{\mathrm{iso}} \, \frac{(1+z)^2}{4 \pi d_L^2} \, Q' \, , \qquad E=\frac{\Gamma}{1+z} \, E' \, . \label{equ:boost}
\end{equation}
Note that for the sake of comparability, we show all neutrino and cosmic ray fluences in the observer's frame, assuming that diffusion, pair production, and photohadronic interactions can be neglected. In \Sec~\ref{sec:cr}, however, we discuss the impact on UHECR observations including pair production and photohadronic losses.
Additionally, we assume that neutrinos are subject to flavor mixing using the mixing angles $\theta_{12} = 0.587$, $\theta_{13} = 0.156$, $\theta_{23} = 0.670$, and $\delta_{\text{CP}} = 1.08\, \pi$, taken from \citet{Fogli:2012ua} for the normal (mass) hierarchy.
For reference, it will be also illustrative to show the cosmic ray proton fluence if all protons were allowed to escape over $t_{\text{dyn}}'$, which represents the maximal possible ``leakage'' from the source. This fluence can be obtained from \equ{boost} using $Q_p'=N_p'/t_{\text{dyn}}'$.

There are three important features of our approach. First of all, we relate everything to the prompt phase, which is implied by using the flux during that phase in \equ{eisobol}. 
The proper densities in \equ{norm} describe the (steady) proton and photon densities in that phase. 
We do not specify the origin of the target photons, such as synchrotron emission of co-accelerated electrons or inverse Compton scattering. Second, we consider emitted neutrino and cosmic ray fluences instead of fluxes, which implies that we do not need to resolve the time-dependence of the emissions.  For instance, the cosmic ray protons emitted with a different escape mechanism may not be emitted at the same time as the gamma-rays. And, third, we compute the fluences per shell, which may seem a bit peculiar, but has the advantage that our approach can describe dynamical changes of the fireball parameters over time, such as collisions at different radii. If all collisions are alike, as it is often assumed, one can easily obtain the result by summing over $N \simeq T_{90}/t_v$ such collisions. 

\section{Direct escape from an expanding shell}
\label{sec:direct}

In this section, we discuss ``direct escape'' as the UHECR escape mechanism. This escape mechanism refers to the escape of protons without further interactions. Since protons (at lower energies) are magnetically confined, it is clear that only protons from the outer edges of the shells can escape directly. While this contribution may be generically expected to be small, we will demonstrate that it is an energy-dependent fraction of protons which can directly escape, and that the direct escape can dominate over the escape of neutrons produced in photohadronic interactions in regions of the parameter space. 

We describe the direct escape of protons from an expanding shell by the fraction of particles which can escape, relative to the densities in \equ{norm}. Assuming that the particles are isotropically distributed in the shell, the number of escaping particles is proportional to the escape volume. We assume that 
particles can escape from within a shell with thickness $\lambda_{\text{mfp}}'$ without interaction, where $\lambda_{\text{mfp}}'$ refers to the mean free path. The fraction of escaping particles $f_{\mathrm{esc}}$ present in the collision phase can be estimated from the ratio between escape volume and isotropic volume as
\begin{equation}
	f_{\mathrm{esc}} \equiv \frac{V'_{\text{direct}}}{V'_{\text{iso}}} \simeq \frac{1}{2} \cdot \frac{4\pi \, (r^2 + (r - \Delta r')^2) \, \lambda'_{\text{mfp}}}{4\pi \, r^2 \, \Delta r'} \simeq \frac{\lambda'_{\text{mfp}}}{\Delta r'} \quad . \label{equ:vesc}
\end{equation}
Here it is taken into account that there are in fact two escape regions in each shell (inner and outer edge), and that only half of the particles along the mean free path propagate in the outwards direction. The proton fluence from escape can then be computed from \equ{boost} using 
\begin{equation}
Q_p'=\frac{N_p'}{t_{\text{dyn}}'} \cdot f_{\mathrm{esc}}  = \frac{N_p'}{t_{\text{dyn}}'} \cdot \frac{\lambda_{\mathrm{mfp}}'}{\Delta r'} = \frac{N_p'}{t_{\text{eff,dir}}'} 
\, 
\label{equ:effesc}
\end{equation}
with an effective direct escape timescale $t_{\text{eff,dir}}' \equiv t_{\text{dyn}}' \Delta r'/\lambda_\mathrm{mfp}'$.

The relevant question is: what determines  $\lambda_{\text{mfp}}'$ of the protons?  Since we consider GRBs, protons (or other charged particles) will interact with the magnetic field of the plasma. This is most often thought to trap the protons inside the shock due to magnetic confinement, while neutral particles, such as neutrons, are considered to escape. When the interactions with the magnetic field are treated as scattering, the Larmor radius 
\begin{equation}
	 R'_L = \frac{E_p'}{e \, B'}  \simeq 33.3 \, \centi\meter \cdot \left( \frac{E_p'}{\giga\electronvolt} \right) \cdot \left( \frac{10^5 \, \text{G}}{B'} \right) \label{equ:RLprotons}
\end{equation}
can be used to estimate the mean free path.
That means that effectively everything within $R_L'$ of the edges will eventually escape from the shells, which we call ``direct escape''. Using the analogy to scattering, it corresponds to the region where the particles can escape without being scattered at all, and without having lost energy.
Now, $R'_L$ is proportional to $E_p'$, and $R'_L = c t_{\mathrm{acc}}'$ for $\eta=1$ (see \equ{tacc}). Therefore, it is easy to see that all protons will directly escape at the highest energy if $\eta=1$ and the maximal proton energy is limited by the dynamical timescale, where $c t_{\mathrm{acc}}'=\Delta r'$. This is the region where direct escape of UHECR  dominates. For lower acceleration efficiencies, or synchrotron- or photohadronic-loss limited maximal proton energies, the direct escape component will be smaller, and the question of what kind of escape dominates will be more complicated.

As long as the proton and photon densities are not too high, the mean free path of protons is given by
\begin{equation}
	\lambda'_{\text{mfp}}(E') =  \min \left[ \Delta r', R'_L(E') \right] \, , \label{equ:mfp}
\end{equation}
This can be used in \equ{vesc}, where the first term ensures that $f_{\mathrm{esc}} \le 1$. Note however that this only applies as long as the source is optically thin. We will discuss the optically thick regime, in which the photohadronic interactions dominate, in the next section in greater detail, since it turns out that it is different from the direct-escape dominated region.

After having identified the different processes which limit the mean free path of the protons, we still need to consider how the expansion of the shell affects the scattering treatment of the particles. Beyond the collision zone, the shell width expands as $\Delta r' \propto r  \, (\propto t)$. As a consequence, the volume of the shell grows as $V' \propto r^3$. This expansion is generally taken to be adiabatic since at this stage the photons are assumed to no longer couple to the electrons (or protons). The general relation between energy and volume of a gas is given by
\begin{equation}
	E' \propto  V'^{-(\hat \gamma-1)} \propto \,  r^{-3 (\hat \gamma-1)}  \, , \label{eq:Escaling}
\end{equation}
for the adiabatic index $\hat \gamma$. Since we know the scaling of the volume as well as the total energy, we can also derive the scaling of the magnetic field inside the plasma as
\begin{equation}
	B' \propto \sqrt{\frac{E'}{V'}} \propto r^{-\frac{3}{2} \hat \gamma} \quad .  \label{eq:Bscaling}
\end{equation}
If we assume that the energy of a single particle in the plasma scales in the same way as the total energy of the plasma (see Eq.~\eqref{eq:Escaling}), the Larmor radius $R_L'$ of the particles changes with time (or radius $r$) as
\begin{eqnarray}
	R_L' &\propto \frac{E'}{B'}& \propto r^{-3 \left( \frac{\hat \gamma}{2}-1 \right)} \, . \label{equ:Larmorscaling}
\end{eqnarray}
The usual assumption for a relativistic gas (or plasma) is $\hat \gamma =4/3$, which leads to $R_L' \propto r$.
This scaling of the Larmor radius has an interesting consequence: the ratio between $R_L'$ and the width of the shell $\Delta r' \propto r$ is constant.
By computing neutrino and proton fluences (instead of fluxes), we do not need to identify when the particles actually escape, we just compute the fraction of escaping particles $f_{\mathrm{esc}}$. 
This means that we can evaluate \equ{vesc} in the collision phase using $r \simeq r_C$ with the corresponding proton density. Note that  other possible values for $\hat \gamma$ are $\hat \gamma = 5/3$ (mono-atomic classic gas), $7/5$ (diatomic classic gas), or $9/7$ (three-atomic gas with non-static bindings).
From \equ{Larmorscaling}, one can easily see that $\hat \gamma = 4/3$ represents a special case that simplifies the scaling and just gives stability between losses and escape. 
For $\hat \gamma > 4/3$, the adiabatic energy loss is faster than the escape, and the particles are more strongly confined for larger radii.  In that case, our computations may over-estimate the direct escape, and most particles will directly escape early.
For $\hat \gamma < 4/3$, the Larmor radius increases faster than the particles lose energy, and the particles are getting less confined at larger radii. With larger radii, particles with lower and lower energies can finally escape. In that case, especially lower energy protons may escape in later phases.

\begin{figure*}[tp]
	\centering
	\includegraphics[width=0.8\textwidth]{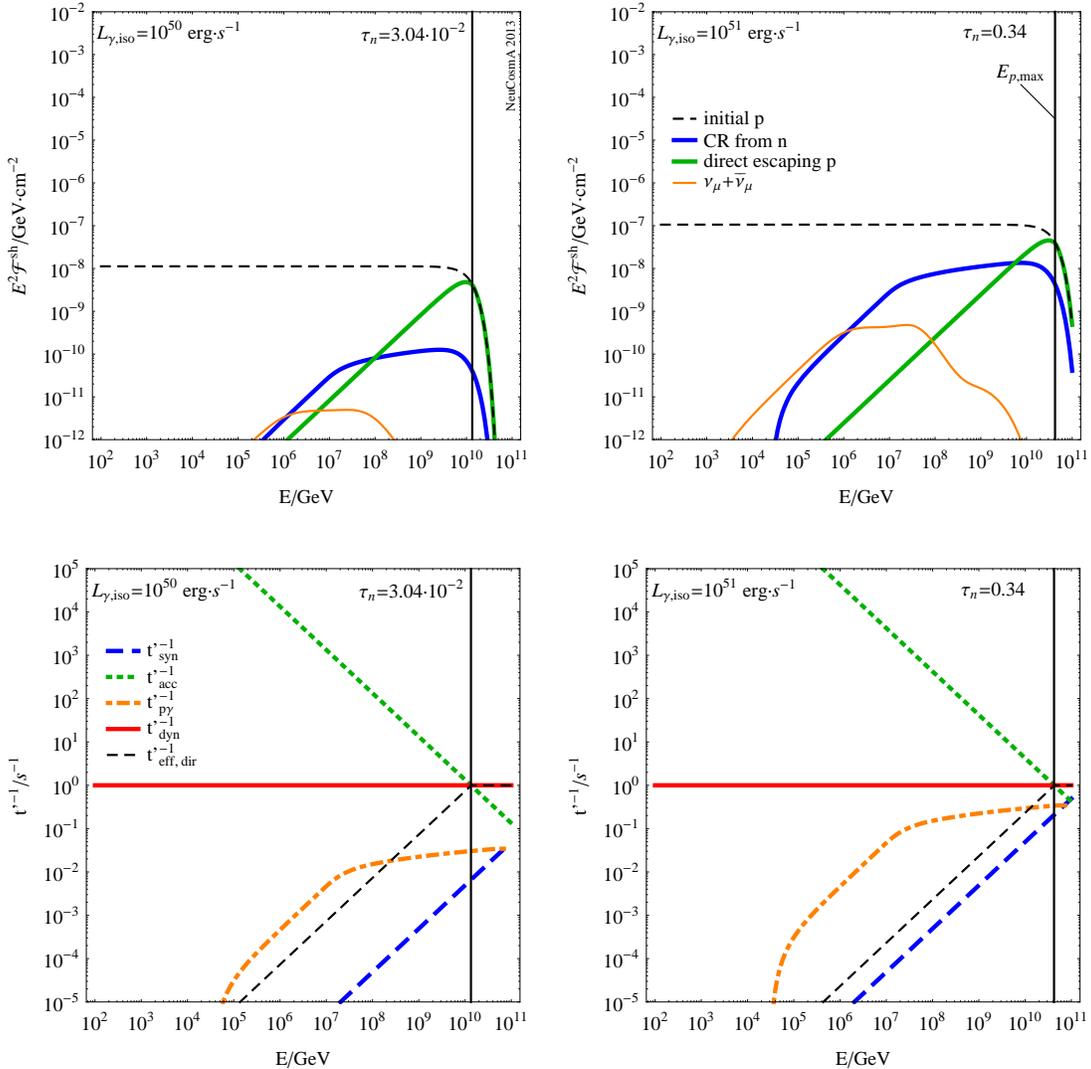}
	\caption{\label{fig:optthin} Particle fluences per shell (upper row) and inverse timescales of different processes (in SRF, lower row) as a function of $E$ in the observer's frame. The figure shows two different parameter values for $L_{\gamma,\mathrm{iso}}$ in the different columns, where the other burst parameters are fixed to  $\Gamma = 300$, $t_v = 0.01 \, \second$, 
 $\eta = 1$,  $\epsilon_e/\epsilon_B = 1$,  $f_e = 0.1$, $\alpha_\gamma = 1$, $\beta_\gamma = 2$,  $\varepsilon'_{\gamma,b} = 1 \, \kilo\electronvolt$, and $z = 2$. Both examples are for the optically thin (to neutron escape) case, where the optical thickness $\tau_n$ is given in the panels. For the cosmic rays, only adiabatic energy losses are taken into account for the propagation. See main text for details. }
\end{figure*} 

Two examples of direct escape-dominated bursts can be found in the different columns of \figu{optthin} for the parameter sets given in the plot. The upper row shows the particle fluences, where ``initial p'' stands for the case if all protons were able to escape over the dynamical timescale, ``CR from n'' represents the cosmic rays through neutron escape from photohadronic interactions, ``direct escaping p'' for direct proton escape, and $\nu_\mu + \bar\nu_\mu$ for the muon neutrino fluence including flavor mixing.\footnote{Note that for the sake of comparability, the CR spectra shown here are ``at the observer'' assuming that the CRs receive the same boost and losses as the neutrinos. In particular, the spectra here are without any losses during propagation apart from the adiabatic losses due to the cosmic expansion.} In the lower panels, the acceleration timescale and the considered energy loss/escape timescales are shown, where the direct escape effective timescale is defined in \equ{effesc}. It can be read off from these panels 
that in both cases the dynamical timescale limits the maximal proton energy. In the upper left panel, the direct escape clearly dominates. Since the acceleration efficiency $\eta=1$, practically all protons escape at the highest energy, which is where the dynamical timescale, acceleration timescale, and direct escape timescale meet.
 However, in that case hardly any neutrinos are produced due to the low photohadronic interaction rate. The upper right panel represents the typical case for the optically thin (to neutron escape) source, where both  substantial neutrino and neutron fluxes are produced. The additional component from direct escape still dominates at the highest energies, while for energies below $10^{9.5} \, \giga\electronvolt$, the neutron flux dominates the cosmic ray production. This can be also read off from the corresponding timescales in the lower panel.
Note that direct escape strongly depends on the acceleration efficiency: if $\eta \ll 1$, the Larmor radius will be much smaller than $\Delta r'$ at the maximal proton energy, and the direct component becomes suppressed. 

Note that the results in this section can be only interpreted as rough estimates, and there may be additional escape components compared to the ones discussed here, \eg, diffusion may play a role. We discuss the possible impact of diffusion in \App~\ref{app:diffusion}, where we demonstrate that it does not affect our qualitative conclusions. However, we also point out that a dedicated treatment of diffusion requires a model-dependent solution of the transport equations, which goes beyond the scope of this study, whereas direct escape can be regarded as a guaranteed contribution to the cosmic ray injection.

\section{Optically thick (to proton and neutron escape) case}
\label{sec:thick}

We define the optical thickness to neutron escape as
\begin{equation}
	\tau_n \equiv \left. \frac{t'^{-1}_{p\gamma}}{t'^{-1}_{\text{dyn}}} \right|_{E_{p,\text{max}}} \label{equ:optthicknessn} 
\end{equation}
at the maximal proton energy. Thus, if $\tau_n \gtrsim 1$, neutrons at the maximal proton energy will rather interact than escape, and will therefore be confined. Since $t_{p \gamma}'^{-1}$ increases with energy (see, \eg, \figu{optthin}, lower panels), this optical thickness is typically at its maximum at the maximal proton energy. That is, it applies to the UHECR part of the emission, whereas neutrons at lower energies may escape more easily. Of course, not only the neutrons will interact rather than leave the region, but also the protons. 

For a numerical description, we follow the same approach as in the previous section. We assume that only a fraction of the produced neutrons can directly escape, \ie, we apply the same mechanism to (direct) neutron escape and multiply the neutron injection $Q'_n$ by \equ{vesc}.  The mean free paths for protons and neutrons, respectively, are in the optically thick regime given by
\begin{eqnarray}
	\lambda'_{p,\text{mfp}}(E') & = &  \min \left[ \Delta r', R'_L(E'), c\, t'_{p\gamma} (E') \right] \, ,  \nonumber \\
\lambda'_{n,\text{mfp}}(E') & = &  \min \left[ \Delta r',  c\, t'_{p\gamma} (E') \right] \, . 
\label{equ:mfpall}
\end{eqnarray}

\begin{figure*}[tp]
	\centering
	\includegraphics[width=0.8\textwidth]{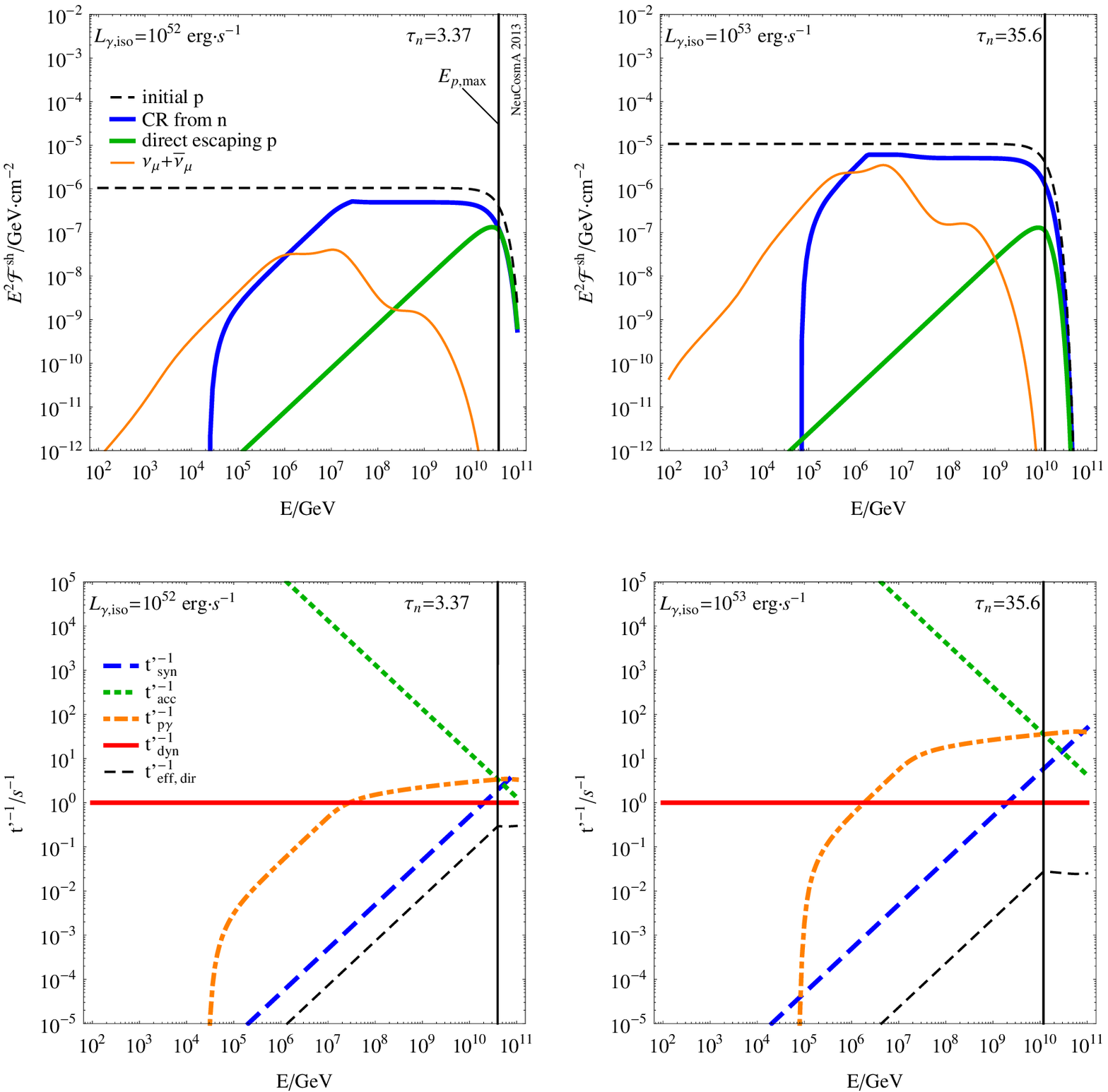}
	\caption{\label{fig:optthick} Particle fluences per shell (upper row) and inverse timescales of different processes (in SRF, lower row) as a function of $E$ in the observer's frame. The figure shows two different parameter values for $L_{\gamma,\mathrm{iso}}$ in the different columns, where the other burst parameters are fixed to the same values as in \figu{optthin}. Both examples are for the optically thick (to neutron escape) case, where the optical thickness $\tau_n$ is given in the panels. See main text for details.}
\end{figure*} 

We show two examples for the optically thick case in the columns of \figu{optthick}. In both cases (see lower panels), the photohadronic interaction rate exceeds the dynamical escape rate by a factor of $\tau_n$ at the highest energies. It also limits the maximal proton energy in both cases. The neutron production is therefore very efficient; see upper panels. However, only the neutrons from the edges can escape, which implies that the dashed curves (corresponding to the escape of all protons over the dynamical timescale) cannot be exceeded, and a level of about 50\% of the dashed curves is reached (since the baryon system contains about 50\% protons and neutrons each in the optically thick limit).
In contrast, the neutrinos from interactions everywhere within the shell can escape, which means that the neutrino fluence becomes relatively enhanced, and the ``one neutrino per cosmic ray'' paradigm does not hold anymore. This can be especially seen in the upper right panel. Note however that the neutrino fluence is typically lower than the cosmic ray fluence, because the neutrons obtain a higher fraction of energy in the interaction.
In none of the discussed cases the direct escape of protons substantially contributes, which is characteristic for the optically thick regime. 

Our approach has several limitations. First of all, one may argue that the neutrons trapped by photohadronic interactions may escape later in a relativistically expanding fireball. Indeed, since the photon density drops as $r^{-3}$, the injection of neutrons ceases and the confinement by photohadronic interactions will come to an end at a certain radius (the ``neutronsphere''), and all remaining neutrons may escape. The level of the neutron density is at the level of the proton density at the end of the confinement phase, which means that a neutron fluence up to the level of the dashed curves in \figu{optthick} might be reached. Therefore, we expect that our ``CR from n'' curves in \figu{optthick} represent a more sophisticated time-dependent calculation within a factor of two. Furthermore, the assumed energy partition fractions may be different than the ones assumed in the optically thick regime.  We also do not consider the effects of muon re-acceleration~\citep{Koers:2007je,Murase:2011cx,Klein:2012ug}, or interactions of pions and kaons~\citep{Kachelriess:2007tr}, which have however much smaller interaction rates than the protons.

There are several subtleties in the optically thick case, which are best illustrated with the pion production efficiency $f_\pi$ relevant for neutrino production. As we show in detail analytically in \App~\ref{app:pion}, these lead to an underestimation of the neutrino production in the optically thick case if the current IceCube method for the computation of $f_\pi$ from \citet{Abbasi:2009ig} is used, which is the foundation for all state-of-the-art GRB stacking analyses, as in \citet{Abbasi:2012zw}. In fact, it turns out that the original formula for $f_\pi$ from \citet{Guetta:2003wi} also applies to the optically thick case if the energy partition is defined with respect to the particle densities within the source, even though it was not derived for that limit. Current state-of-the-art numerical predictions, such as \citet{Hummer:2011ms}, take this into account automatically.

\section{Parameter space study of the cosmic ray-neutrino connection}
\label{sec:space}

For the cosmic ray-neutrino connection, we identify three different regimes:
\begin{description}
\item[Optically thin to neutron escape regime] This is the usual scenario discussed in the literature: the cosmic rays are produced as neutrons and can escape the source (``neutron model''). Additional escape components are negligible, and the ``one (muon) neutrino per cosmic ray'' paradigm applies.
\item[Direct escape regime] Here the cosmic rays from direct escape dominate at least at the highest energy. Since the neutron production by photohadronic processes is sub-dominant, the one neutrino per cosmic ray relationship does not hold, and more cosmic rays than neutrinos will be produced. See \Sec~\ref{sec:direct}.
\item[Optically thick to neutron escape regime] Here the protons and neutrons interact multiple times, and only protons and neutrons on the outer edges of the shells can (directly) escape. The neutrinos, however, can escape from everywhere within the shell, which leads to more neutrinos per cosmic ray than in the optically thin case.
See \Sec~\ref{sec:thick}.
\end{description}
In this section, we discuss the type of cosmic ray source as a function of the GRB parameters. We distinguish the dominant effect by using figures such as \figu{optthin} and \figu{optthick}: if the fluence maximum in the spectrum comes from directly escaping protons, we assign the direct escape category, otherwise the optically thin regime. 
The optically thin and thick cases are distinguished by the optical thickness $\tau_n$, as defined in \equ{optthicknessn}, being smaller or larger than one, respectively. It turns out that either of these three categories can be uniquely assigned (neglecting minor overlap). 

\begin{figure*}[t!]
	\centering
	\includegraphics[width=0.8\textwidth]{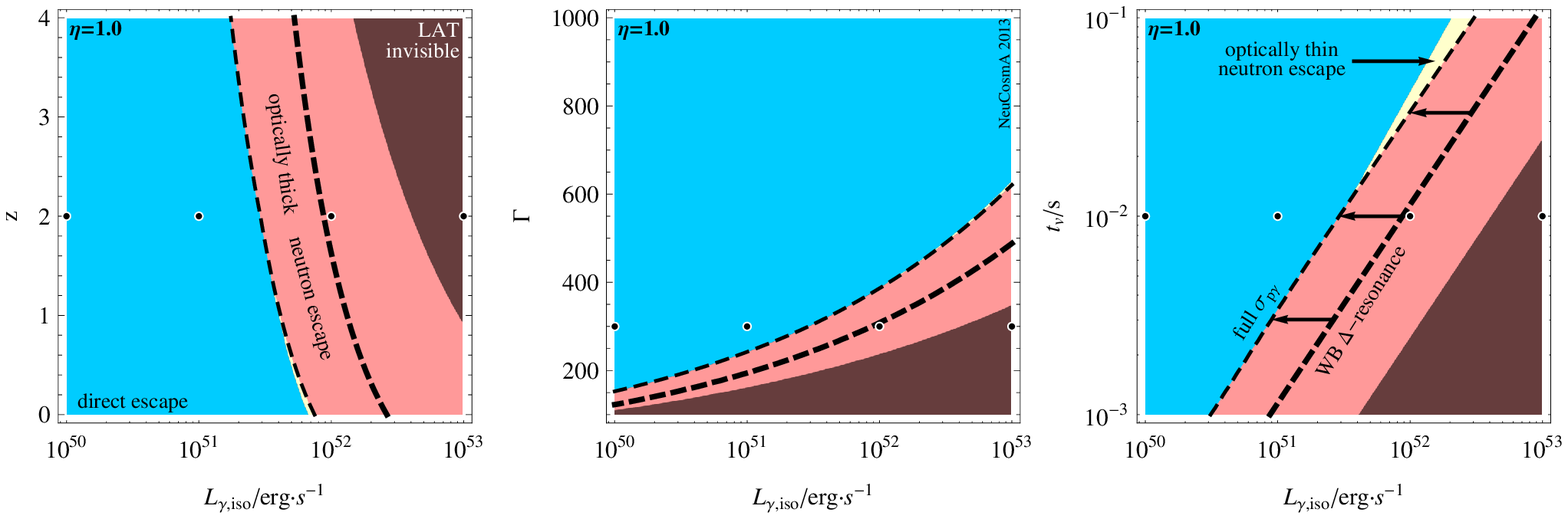}

\vspace*{0.5cm}

	\includegraphics[width=0.8\textwidth]{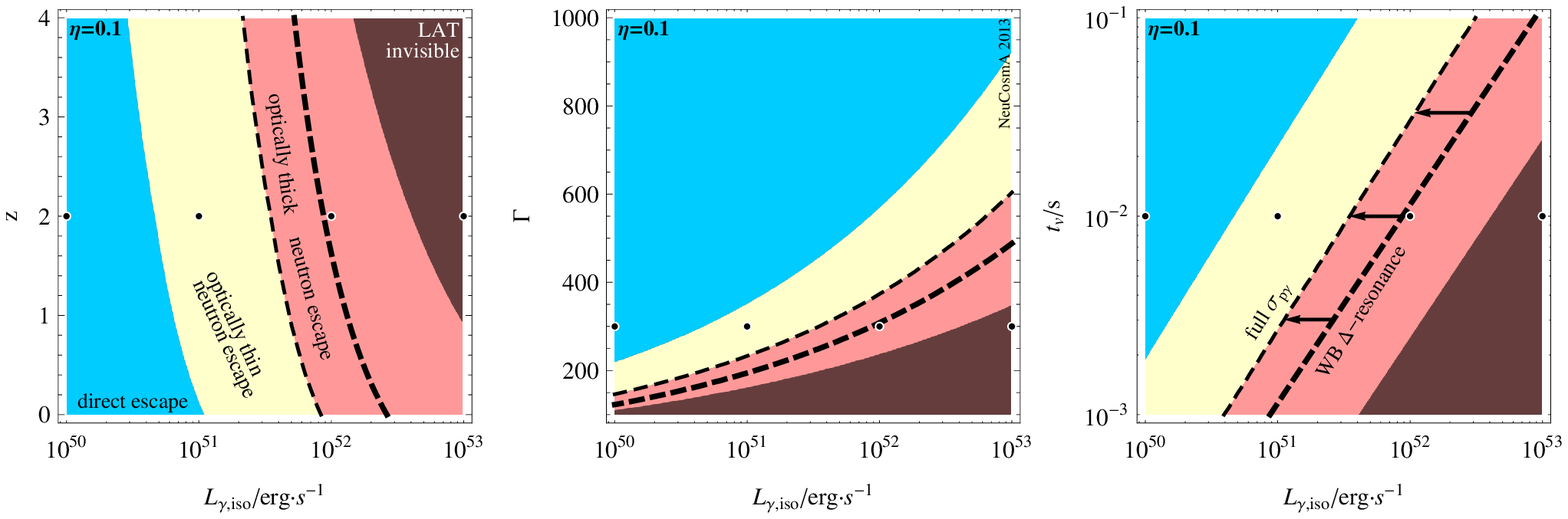}
	\caption{\label{fig:scan}
Parameter space scan of the GRB parameters for classification of regimes. The upper row corresponds to the acceleration efficiency $\eta=1$, the lower row to $\eta=0.1$. 
For the standard parameter values, see caption of \figu{optthin}. Dashed curves mark the limit between optically thin and thick regimes. The thick dashed curve represents the analytical result if the photohadronic interaction rate based on \eq~(3) of \citet{Waxman:1997ti} is used. In the dark-shaded regions ``LAT invisible'', gamma-rays above 30~MeV cannot leave the source anymore due to pair production. That is, sources left of these regions are in principle visible in the full {\it Fermi}-GBM range and may be even observable in LAT, whereas sources within these regions will not exhibit emission into the LAT range. The dots correspond to the parameter sets chosen in \figu{optthin} and \figu{optthick}.}
\end{figure*}

We show a (numerical) parameter space scan in \figu{scan}, where we always plot $L_{\gamma,\mathrm{iso}}$ on the horizontal axis. Let us focus on moderate acceleration efficiencies $\eta=0.1$ first (lower row), which clearly exhibit all three regimes. The optically thin case can be found close to the often used standard parameter values. In this case, the direct escape contribution cannot reach the same level as shown in \figu{optthin} since the maximal achievable proton energy is lower, and therefore the Larmor radius at the maximal energy cannot reach the shell width.  Therefore, the escape of neutrons produced in photohadronic interactions dominates. On the other hand, in the light red (gray) regions, $\tau_n > 1$, which means that the neutrino production is enhanced. In the blue (darker) regions, the direct escape component exceeds the neutron escape at the highest energies. For efficient acceleration, $\eta=1$, see upper row. Here the optically thin to neutron escape region almost vanishes, 
which is due to an enhancement of the direct escape. Remember that for $\eta=1$, all protons can directly escape at the highest energy if the maximal energy is limited by the dynamical timescale.  To summarize, for efficient proton acceleration, the standard case (one neutrino per cosmic ray) only applies in a very small region of the parameter space, and either fewer or more neutrinos per cosmic ray are produced, depending on the parameters. 

In order to better understand the relationship to earlier works, consider the $\Delta$-resonance parameterization proposed by \citet{Waxman:1997ti} (see Eq.~(3), increased by a factor of two because we consider the photohadronic interaction rate, not the pion production rate).  We show in \figu{scan} the separator between optically thin and thick regimes as thick dashed curves for this cross section. Obviously, in all cases the optically thin region reappears and gets enlarged. One reason is that the full numerical computation includes high-energy processes, such as multi-pion production, which enhance the interaction rate at high energies by a factor of a few, and which is not included in the shown $\Delta$-resonance approximation. As a result, photohadronic processes become more important in the numerical result. 

Let us now relate the parameter space to \textit{Fermi}-LAT observations. We hence show in \figu{scan} the ``LAT invisible'' regions, where gamma-rays above $30\,\mega\electronvolt$ cannot leave the source anymore because they exceed the pair production threshold.
That is, sources left of these regions should be visible in the full \textit{Fermi}-GBM range and may be even observable in LAT, while sources in the ``LAT invisible'' regions will not exhibit emission into the LAT range. One can clearly see that the parameter set corresponding to the rightmost dot, associated to the right column of \figu{optthick}, cannot be seen in LAT. 
In addition, even though the optical thicknesses $\tau_n$ (for neutrons) and $\tau_{\gamma\gamma}$ (for photons) are roughly proportional, they still have slightly different parameter dependencies.\footnote{This is mainly due to one important difference in the calculations, namely which frame is considered to be relevant. The maximal proton/neutron energy is calculated in the SRF during our calculation, however the calculation of $\tau_{\gamma\gamma}$ is done for an observed photon energy. Hence there is a difference in redshift $z$ and Lorentz factor $\Gamma$. Moreover, the break energy is important for the calculation of the (energy) densities, but it is not relevant for the optical thickness of the observed gamma-rays, as this calculation refers to photons far above the break.}
Note that, in practice, the LAT emission lasts longer (see, \eg, \citet{Lange:new}), which means that it may come from larger emission radii. 
That is, in a realistic time-dependent model, the fireball may follow a trajectory in the considered parameter space, and may actually visit more than one regime during the burst duration. It remains to be seen what the consequences of such a time-dependent model on the neutrino fluxes will be.

Since \figu{scan} only shows sections through the parameter space, it is instructive to at least have approximate analytical expressions for the different regimes. The interface between the direct escape and optically thin regions can be obtained from the maximal proton energy: if it is dominated by the dynamical timescale, direct escape will dominate at the highest energies since all protons can escape (for $\eta=1$); if it is dominated by synchrotron losses, neutron escape will take over. Taking into account $\eta$, one can estimate that 
\begin{eqnarray}
	&& L_{\gamma,\text{iso}}^{\text{direct}} \lesssim 3.6 \cdot 10^{51} \, \text{erg}\,\reciprocal\second \cdot \eta^{\frac{2}{3}} \cdot \left(\frac{\Gamma}{10^{2.5}} \right)^{\frac{14}{3}} \nonumber \\
        && \hspace{1.5cm} \cdot \left(\frac{t_v}{0.01 \, \second} \right)^{\frac{2}{3}} \cdot \left( \frac{1+z}{3} \right)^{-\frac{2}{3}}   
           \cdot \left(\frac{\epsilon_B}{\epsilon_e} \right)^{-1} \label{equ:borderadsyn}
\end{eqnarray}
limits the direct escape dominated region. The interface between the optically thin and thick regimes (analytical thick dashed curves) can be obtained from $\tau_n \simeq 1$ in \equ{optthicknessn}, using the analytical expression for $t'^{-1}_{p\gamma}$ from \citet{Waxman:1997ti}. The optically thick regime is then estimated  as
\begin{eqnarray}
 	&& L_{\gamma,\text{iso}}^{\text{opt. thick}} \gtrsim 1.1 \cdot 10^{52} \, \text{erg}\,\reciprocal\second \cdot \left( \frac{\Gamma}{10^{2.5}} \right)^5 \cdot \left( \frac{t_v}{0.01 \, \second} \right) 
           \nonumber \\
        && \hspace{2cm} \cdot \left( \frac{\varepsilon'_{\gamma, \text{break}}}{1 \, \kilo\electronvolt} \right)
           \cdot \left( \frac{1+z}{3} \right)^{-1}  \, .\label{equ:borderpgammadamping}
\end{eqnarray}
These formulas allow to estimate how a specific burst can be classified for arbitrary parameters. There are, however, some limitations. First of all, \equ{borderpgammadamping} underestimates the photohadronic interactions, as discussed above. And, second, some of the (numerical) parameter dependencies in \equ{borderadsyn} cannot be reproduced within these assumptions.

\section{Application to specific GRBs}
\label{sec:spec}

\begin{table*}[t]
 \centering
 \begin{tabular}{|l|c|c|c|c|}
  \hline
                                         & SB                & GRB080916C          & GRB090902B          & GRB091024           \\
  \hline
  $\alpha_\gamma$                        & 1                 & 0.91                & 0.61                & 1.01                \\
  $\beta_\gamma$                         & 2                 & 2.08                & 3.80                & 2.17                \\
  $\epsilon_{\gamma,\text{break}}$ [MeV] & 1.556             & 0.167               & 0.613               & 0.081               \\
  $\Gamma$                               & $10^{2.5}$        & 1090                & 1000                & 195                 \\
  $t_v$ [s]                              & 0.0045            & 0.1                 & 0.053               & 0.032               \\
  $T_{90}$ [s]                           & 30                & 66                  & 22                  & 196                 \\
  $z$                                    & 2                 & 4.35                & 1.822               & 1.09                \\
  $S_{\text{bol}}$ [erg cm$^{-2}$]   & $1 \cdot 10^{-5}$ & $1.6 \cdot 10^{-4}$ & $3.3 \cdot 10^{-4}$ & $5.1 \cdot 10^{-5}$ \\
  $L_{\gamma,\text{iso}}$ [erg s$^{-1}$]   & $10^{52}$         & $4.9 \cdot 10^{53}$ & $3.6 \cdot 10^{53}$ & $1.7 \cdot 10^{51}$ \\
  \hline
 \end{tabular}
 \caption{\label{tab:TblA} Properties of four bursts discussed in \Sec~\ref{sec:spec}; see \citet{Baerwald:2010fk} for SB (``Standard Burst'', similar to \citet{Waxman:1997ti,Waxman:1998yy}), \citet{Nava:2010ig} and \citet{Greiner:2009pm} for GRB080916C, \citet{Nava:2010ig} and \citet{Abdo:2009pg} for GRB090902B, and \citet{Nava:2010ig} and \citet{Gruber:2011gu} for GRB091024. The luminosity is calculated with $L_{\gamma,\text{iso}}=4\pi d_L^2\cdot S_\text{bol}/T_{90}$, with $S_{\text{bol}}$ the fluence in the (bolometrically adjusted) energy range $1\,\text{keV}-10\,\text{MeV}$. Adopted from \citet{PhDHummer}.}
\end{table*}

In the previous sections, we have discussed the dependence of the cosmic ray escape on the theoretical parameters, such as $L_{\gamma,\mathrm{iso}}$. However, actually the gamma-ray flux or fluence are the observables, and $L_{\gamma,\mathrm{iso}}$ is just a function of these observables. In addition, for many bursts, the other necessary ingredients, such as redshift and time variability, have been measured. We therefore study in this section the different cosmic ray escape mechanisms for specific bursts; see \Tab~\ref{tab:TblA}. In this case, we use the bolometric fluence as observable, and assume that it is obtained from $N=T_{90}/t_v$ identical collisions, in consistency with the approaches used in \citet{Abbasi:2011qc,Abbasi:2012zw}, and \citet{Hummer:2011ms}. 

The standard burst ``SB'' has been inspired to produce a spectrum similar to \citet{Waxman:1997ti,Waxman:1998yy}. The other three bursts have been actually observed, and their properties can be taken from the literature (see table caption). 
GRB080916C is one of the brightest bursts ever seen, although at a large redshift, and one of the best studied {\it Fermi}-LAT bursts. The gamma-ray spectrum of GRB090902B has a very steep photon spectral cutoff and a smaller redshift, although $\Gamma$ is very high as well. GRB091024 can be regarded as a typical example representative of many {\it Fermi}-GBM bursts~\citep{Nava:2010ig}, except for the long duration.
Note that the neutrino spectra from these GRBs have been also discussed in \citet{Winter:2012xq} and \citet{Baerwald:2012kc}.

\begin{figure*}[tp]
	\centering
	\includegraphics[width=0.8\textwidth]{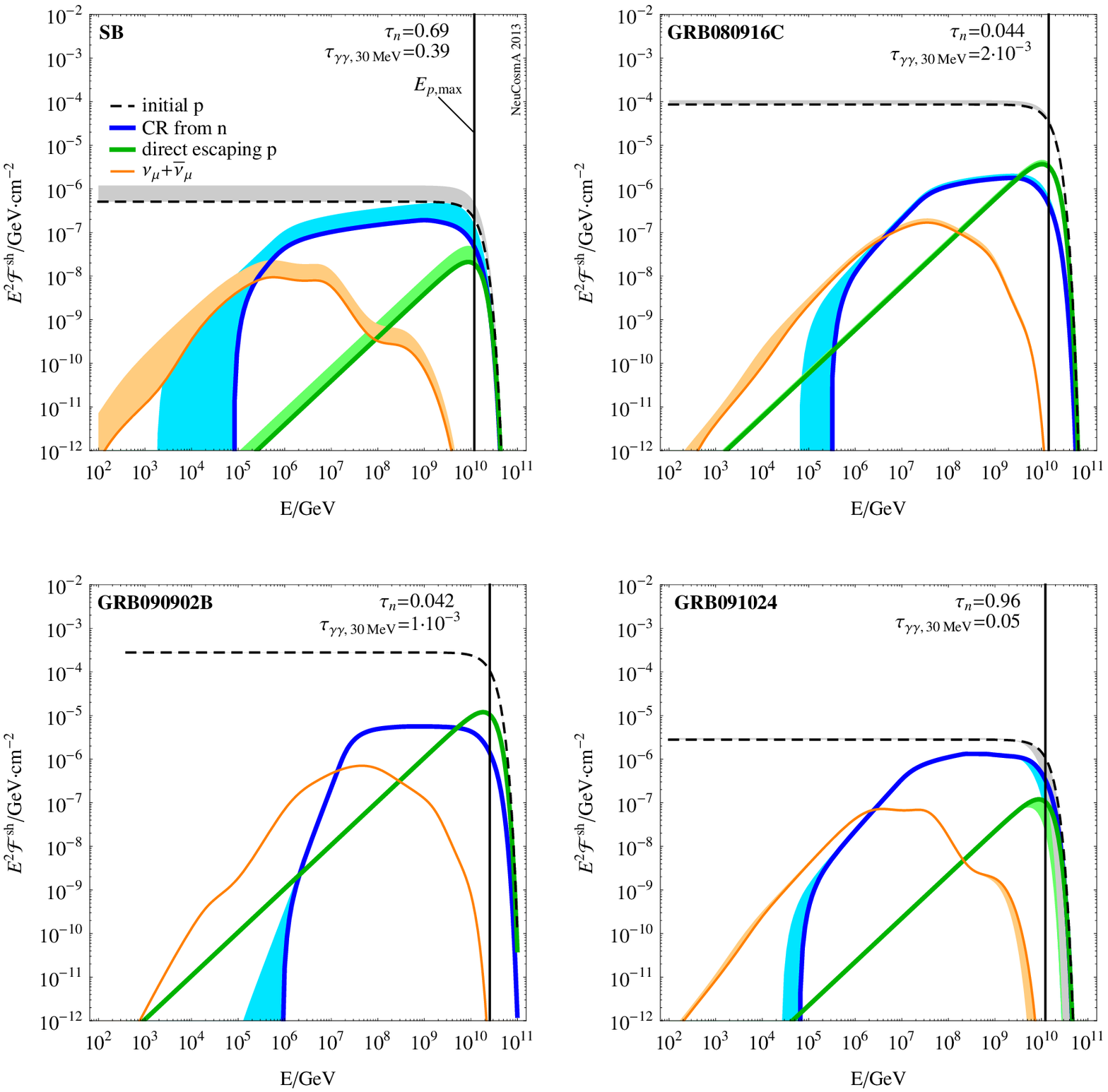}
	\caption{\label{fig:real}The expected cosmic ray  and neutrino fluences for the GRBs listed in \Tab~\ref{tab:TblA}. The maximal proton energy and the optical thickness are given/marked in the different panels. The thick curves correspond to the photon fluence and energy range given in the caption of \Tab~\ref{tab:TblA}, whereas the shadings represent a bolometric correction beyond this range; see main text. In each panel, we also give the optical thickness to neutron and photon escape  (the latter for $30 \, \mega\electronvolt$ gamma-rays). Note that we use an acceleration efficiency $\eta = 0.1$ for these simulations.
}
\end{figure*}

We show in \figu{real} the expected cosmic ray  and neutrino fluences for the GRBs listed in \Tab~\ref{tab:TblA} (thick curves). In each panel, we also give the optical thickness to neutron escape and photon escape (for $30 \, \mega\electronvolt$ gamma-rays). Note that we use an acceleration efficiency $\eta = 0.1$ for these simulations; a higher acceleration efficiency would pronounce the direct escape component. The burst SB is a typical example for an optically thin burst with a substantial amount of neutrino production. The contribution of the direct escape component depends on the proton acceleration efficiency, and in this case it is suppressed. Therefore, the relationship one neutrino per cosmic ray holds. On the other hand, the two high-$\Gamma$ {\it Fermi}-LAT bursts GRB080916C and GRB090902B exhibit a clear direct escape domination, which is a feature of the large $\Gamma$; see \figu{scan} (second row, second column).  It is clear that GRBs will be only 
observed in LAT if they are left of the dark-shaded regions in \figu{scan}. An even larger value of the photon energy, such as 100~MeV, will extend these regions further to the left, which means that the LAT-associated parameter space region tends to overlap the direct escape regime. Therefore, LAT-observed GRBs tend to directly emit UHECR protons at the highest energies, at least in the LAT emission phase. The situation is different for GRB091024 in the lower right panel of \figu{real}, which is on the edge of the optically thick regime. For this burst, again cosmic rays from escaped neutrons dominate, and the neutrino production follows the standard assumption.

In order to illustrate the impact of the minimal and maximal photon energies in \equ{targetphoton}, we illustrate the impact of a ``bolometric correction'' as shaded areas in \figu{real}. This bolometric correction takes into account that the gamma-ray fluence has only been observed in a certain energy range, whereas it cannot be excluded that lower and higher energy photons are present in the source as well, either because they are outside the detection energy range, or because they cannot escape from the source.\footnote{
For this correction, we fix the gamma-ray spectrum in the observed energy range from the observed fluence, and then linearly extrapolate the spectrum (on a double log plot) to the range between $0.2 \, \electronvolt$ (SRF) and $100 \, \mega\electronvolt$ (observer's frame). This extended range is motivated by the fact that high energy protons then always find sufficiently many low energy photons as interaction partners, and low energy protons find enough high energy photons.  The $100 \, \mega\electronvolt$ are chosen in the observer's frame since they correspond to a typical {\it Fermi}-LAT energy.} The proton density is then calculated from energy partition using the extended energy range according to \equ{norm}. The bolometric correction increases it in all cases due to photons not accounted for in the observation, and therefore the normalization of the spectra increases, including that of the ``initial protons''. In neither of the cases, the neutrino spectrum is very much affected by this bolometric 
correction, apart from the normalization change.\footnote{This correction depends mostly on the upper spectral index of the photon spectrum. It is $\varepsilon'^{-2}$ in the upper left panel, which leads to a logarithmic dependence on the maximal proton energy, and steeper in the other cases, which leads to a (stronger) power law suppression.} The extension of the photon energy range, however, hardly affects the neutrino spectral shape. On the other hand, the extension of the photon spectrum to higher energies has a significant effect on the neutron spectra at low energies.
In the lower right panel, the bolometric correction even leads to a lower maximal proton energy, which is because photohadronic energy losses take over to limit the maximal proton energy.
Additionally, we checked that the gamma-rays from $\pi^0$-decays produced by the three actually measured bursts are below the LAT bounds or observations for these bursts.

In summary, {\it Fermi}-LAT observed GRBs seem to a have a strong direct escape component of cosmic rays, at least in that emission phase, whereas the direct escape contribution of typical GBM bursts depends on the proton acceleration efficiency. Therefore, conclusions on the cosmic ray-neutrino connection will depend on the actual burst sample including the specific parameters of the fireballs, and the time evolution of the fireball properties, which we did not discuss in this section. A possible bolometric correction beyond the observed energy ranges of the gamma-rays typically has a small effect, as long as $\beta_\gamma \gtrsim 2$.

\section{Impact on UHECR observations}
\label{sec:cr}

So far, we have discussed the cosmic ray and neutrino emission from a single source in terms of its fluence. However, especially for cosmic rays, single sources may not be resolvable, and the injection as a function of redshift determines the observed spectrum and cosmogenic neutrino flux. 
There are four key ingredients which determine the observed cosmic ray spectrum at the highest energies: the injection spectral index in the co-moving frame (corresponding to our $\alpha_p$), the maximal particle energy (our $E_{p, \mathrm{max}}$), the evolution of the density of sources with redshift, and the chemical composition of the UHECRs. Here we focus on protons as cosmic rays and assume the standard GRB evolution (\citet{Hopkins:2006bw} star formation rate including \citet{Kistler:2009mv} correction for GRBs), which leaves $\alpha_p$ and $E_{p, \mathrm{max}}$ as free parameters. For a recent review on cosmic ray transport and models, see \citet{Aloisio:2012ba} and references therein.

\begin{figure*}[t!]
\centering
\includegraphics[width=0.8\textwidth]{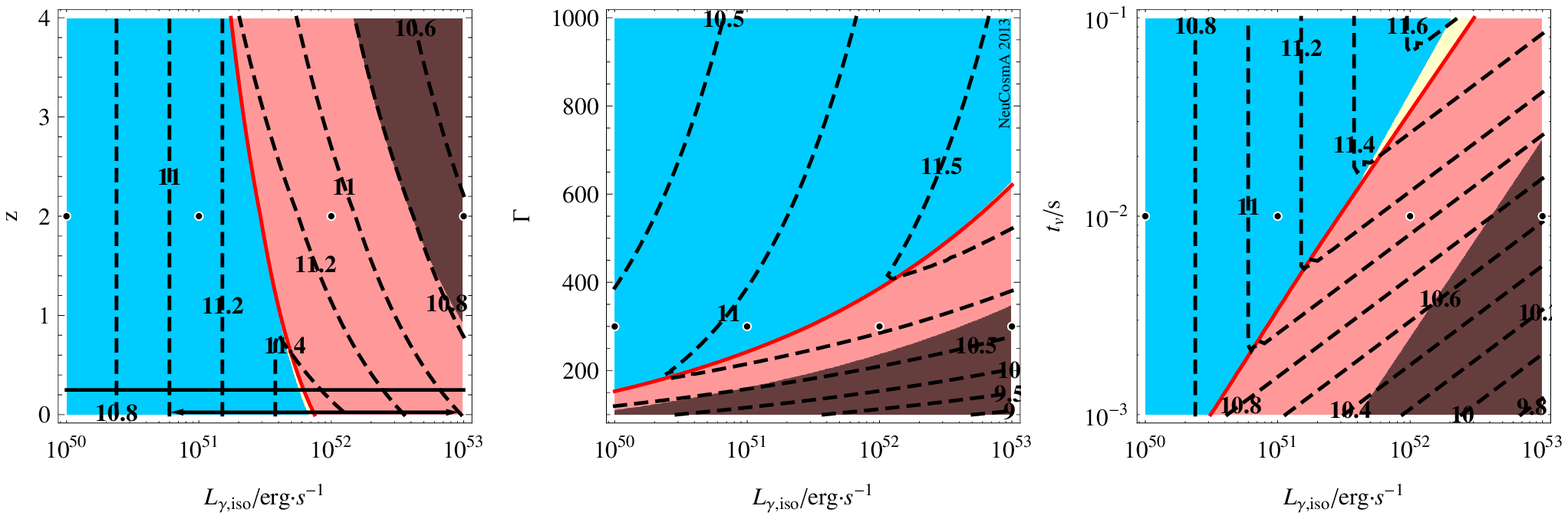}
\includegraphics[width=0.8\textwidth]{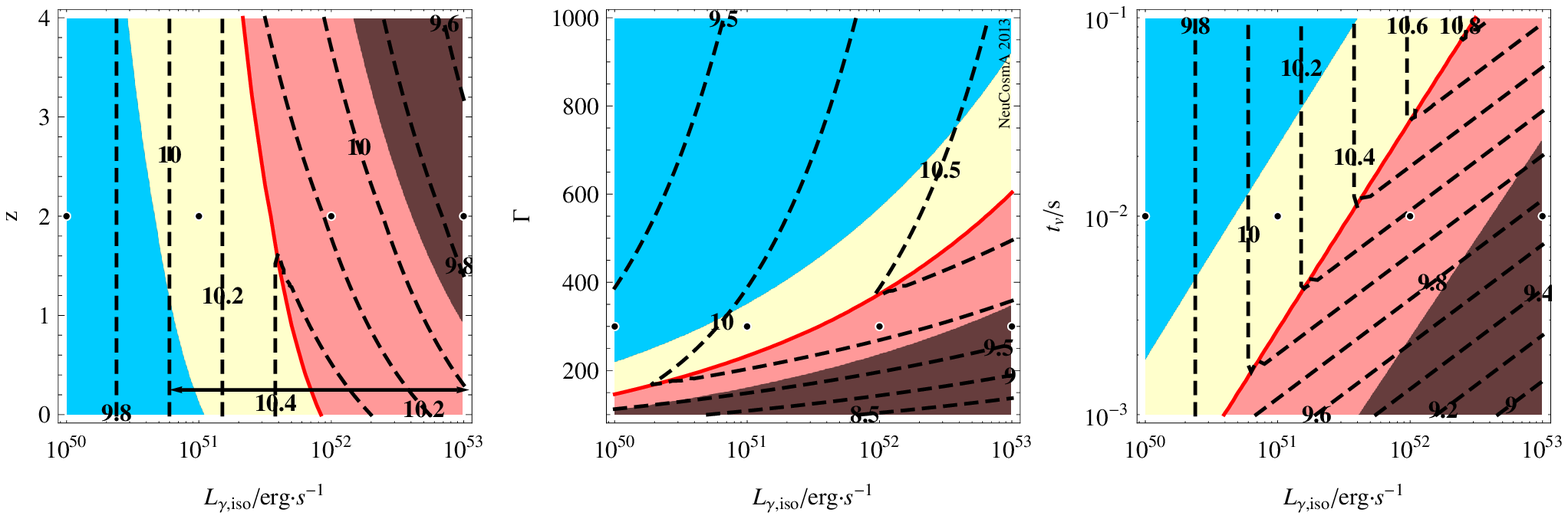}
\caption{\label{fig:epmax} Maximal proton energy $\log_{10} (E_{p,\mathrm{max}}/\giga\electronvolt)$ (contours) in the source (cosmologically co-moving) frame as a function of the GRB parameters. The upper row is for $\eta=1$; the lower row, for $\eta=0.1$. Same regions and parameters as in \figu{scan}.}
\end{figure*}

In order to describe the UHECR spectrum, it is a necessary condition that high enough proton energies can be obtained in the co-moving (source) frame. We show in \figu{epmax} the maximal proton energy $\log_{10} (E_{p,\mathrm{max}}/\giga\electronvolt)$ (contours) in the co-moving frame as a function of the GRB parameters, similar to \figu{scan}. The upper row is shown for an acceleration efficiency $\eta=1$; the lower row, for $\eta=0.1$.
One can clearly see the symmetry around the transition curve between the optically thin (or direct escape dominated) and optically thick cases, where the proton energy is limited by photohadronic losses in the latter case. The highest proton energies can be obtained along this transition curves, but the overall dependence of the maximal proton energy on the model parameters is relatively weak. The main impact comes from the acceleration efficiency (compare upper with lower row): for $\eta=1$, about an order of magnitude higher proton energies can be achieved than for $\eta=0.1$. 

For $E > 10^{10} \, \giga\electronvolt$, the mean free path of the protons is of the order of 1~Gpc ($z \simeq 0.25$), while for $E > 10^{11} \, \giga\electronvolt$, it is of the order of 100~Mpc ($z \simeq 0.024$); the corresponding parameter spaces are illustrated by arrows in the $L_{\gamma,\text{iso}}$-$z$ panels of \figu{epmax}. 
Therefore, for the UHECR spectrum, only low redshifts are relevant, while for the cosmogenic neutrino flux, the full redshift range can contribute (see, \eg, \citet{Kotera:2010yn}), and especially the very highly energetic protons will lead to a substantial neutrino production due to energy losses on the cosmic microwave and infrared backgrounds. From the redshift dependencies  in \figu{epmax} it is clear that for $\eta=1$ (upper row, left plot), all bursts for the chosen parameters can produce the UHECRs above about  $10^{10} \, \giga\electronvolt$; and only bursts between $L_{\gamma,\mathrm{iso}} \simeq 6 \cdot 10^{50} \, \mathrm{erg}\, \reciprocal\second$ and $10^{53} \, \mathrm{erg}\, \reciprocal\second$, above about $10^{11} \, \giga\electronvolt$ (illustrated by arrows). For $\eta=0.1$ (lower row, left plot), the range $6 \cdot 10^{50} \, \mathrm{erg}\,\reciprocal\second \lesssim L_{\gamma,\mathrm{iso}} \lesssim 10^{53} \, \mathrm{erg}\,\reciprocal\second$ can produce the UHECRs above about  $10^{10} \, \giga\electronvolt$, and energies as high as  $10^{11} \, \giga\electronvolt$ are difficult to reach. Note, however, that these parameter ranges are found for the other parameters fixed to their standard values. Nevertheless, if GRBs are the sources of the UHECRs, this discussion illustrates that either the acceleration must be very efficient, or only GRBs from a very narrow parameter space region contribute at the very highest energies. 

\begin{figure*}[t!]
\centering
\includegraphics[width=0.5\textwidth]{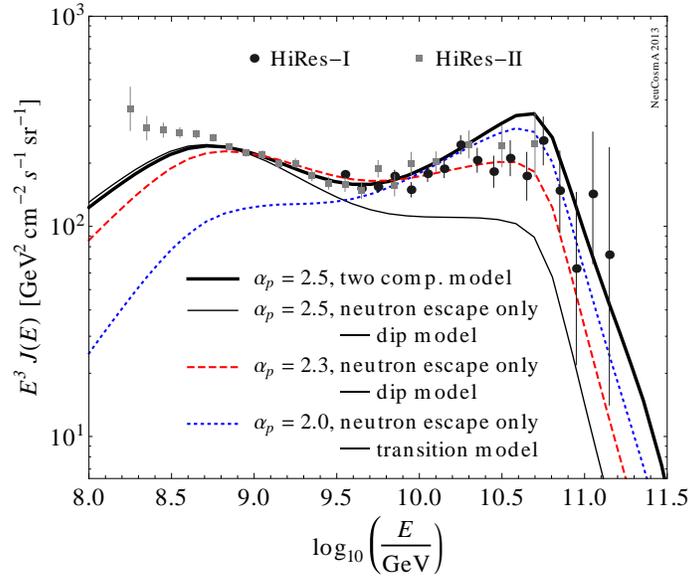}
\caption{\label{fig:twocomp} Example of observed cosmic ray (proton) spectrum for different assumptions at the source and different spectral injection indices $\alpha_p$. Here ``two component escape'' refers to the combined neutron and direct escape components.
The source redshift evolution is assumed to follow the GRB rate (star formation rate from \citet{Hopkins:2006bw} including \citet{Kistler:2009mv} correction for GRB evolution). The HiRes data are taken from \citet{Abbasi:2007sv}. In this example it is assumed that all bursts are alike in the co-moving frame with $\eta=1$, 
$t_v =3.3 \cdot 10^{-3} \, \second$, $\Gamma = 10^{2.5}$, $L_{\gamma,\mathrm{iso}}=7 \cdot 10^{51} \, \mathrm{erg}\cdot\mathrm{s}^{-1}$, $\epsilon'_{\gamma,\mathrm{break}} = 14.76 \, \kilo\electronvolt$, $\alpha = 1$, $\beta = 2$, and $k=1$ in \equ{protons}, where $t_v$ is given in the co-moving frame. The resulting maximal proton energy with these parameters is $1.9 \cdot 10^{11} \, \giga\electronvolt$ (co-moving frame). The normalization of the spectra is arbitrarily chosen, \ie, not based on a fit. }
\end{figure*}

In the following, we adopt a very pragmatic point of view to show the impact of the direct escape on the observed cosmic ray spectrum: we choose a reasonable parameter set for a GRB with a  high enough proton energy from a high acceleration efficiency $\eta=1$ which exhibits the direct escape component at the highest energies, as well as the standard neutron escape component at lower energies, such as in \figu{optthin}, upper right panel. Then we assume that all bursts are alike in the co-moving frame, \ie, we use a universal cosmic ray injection function based on this burst and following the standard GRB evolution, with an arbitrary normalization.  The burst parameters are given in the caption of \figu{twocomp}, and the maximal proton energy is found to be $1.9 \cdot 10^{11} \, \giga\electronvolt$, with which the protons are injected into the interstellar medium. For the cosmic ray transport, we use a deterministic kinetic equation solver assuming continuous energy losses, which is based on \citet{Ahlers:2009rf,Ahlers:2010fw}. After the cosmic ray propagation, the normalization 
of the cosmic spectrum is chosen to reproduce the HiRes data~\citep{Abbasi:2007sv}. Note that the specific weight of the direct versus neutron escape depends on the chosen parameter set, and that a more refined treatment of the leakage may lead to adjustments. However, the idea here is that the presence of an additional escape component can lead to a spectral break in the cosmic ray spectrum from one source, as it is visible in \figu{optthin}, upper right panel. In addition, note that in principle information on the cosmic energy budget can be obtained from this procedure. Since a detailed discussion goes beyond the scope of this study, it will be performed elsewhere.

Now the only remaining free parameter is $\alpha_p$. In that context, two often used approaches for the reproduction of the UHECR shape are: 
\begin{description}
\item[The ``dip model''] using an injection index at the source $\alpha_p \gtrsim 2.5$ (depending on the source evolution). It can reproduce the dip and the ankle very well, \ie, the region above about $10^9 \, \giga\electronvolt$, but the large $\alpha_p$ is difficult to motivate from Fermi shock acceleration. Note that we for now neglect the problems with the energy budget in this model, see \citet{Murase:2008mr} or \citet{Berezinsky:2002nc}.
\item[The ``transition (or ankle) model''] using an injection index at the source $\alpha_p \simeq 2.0$. It can reproduce the CR spectrum above the ankle, \ie, the region above about $10^{10} \, \giga\electronvolt$, but the energy range below the ankle requires the transition to a different (galactic?) component.
\end{description}
Normally, the escape mechanism from the source is not specified, only the cosmic ray injection spectrum (with index $\alpha_p$ and maximal proton energy). We directly compute the cosmic ray direct escape and neutron escape spectra with our source model. Note that $\alpha_p$ is the proton injection spectral index in our model, not the cosmic ray injection index. Since multi-pion processes make the neutron spectrum somewhat harder (see, \eg, \figu{real}), there is a slight difference compared to the usually assumed injection index. Our index $\alpha_p$ refers directly to the acceleration processes in the source.

Here we propose a third possibility: a {\bf two-component escape model} for GRBs with both the neutron and direct escape components. 
The result is illustrated in Fig.~\ref{fig:twocomp} as a black thick solid curve. For $\alpha_p=2.5$, the shape of the cosmic ray spectrum around dip and ankle is almost perfectly reproduced. It is obvious that because of the large $\alpha_p$, the model works better than the transition model with $\alpha_p=2.0$ (blue dotted). The red dashed curve represents a dip model with only the neutron escape ($\alpha_p=2.3$), which obviously reproduces the generic features of the observed spectrum, but not as pronounced as the two-component model. A comparison of the two curves for $\alpha_p=2.5$ (two component versus neutron escape only) illustrates the effect of the second escape component: it enhances the high energy part. The enhancement depends on the chosen parameter set and value of the acceleration efficiency. Note that  Bethe-Heitler losses also contribute to the dip in the two-component model, but the effects of Bethe-Heitler losses and spectral break in the injection spectrum add up. 

We have also tested the impact on the cosmogenic neutrino flux. A two-component model as shown in \figu{twocomp} for $\alpha_p=2.5$ in combination with the GRB strong evolution would lead to a very large production of cosmogenic neutrinos close to the current bound~\citep{NeutrinoTalk}, which should be testable in the very near future. This is a generic feature common to all models with strong source evolution and a pronounced rise after the ankle in the cosmic ray spectrum; see, \eg, \citet{Kotera:2010yn}.  

\section{Summary and conclusions}
\label{sec:summary}

We have discussed the escape of UHECR protons from GRBs in the internal collision (prompt) phase  in the framework of the GRB fireball model. We have identified three different regimes in terms of the dominating UHECR injection mechanism: 
\begin{enumerate}
\item {\bf Optically thin to neutron escape regime.} Neutrons from photohadronic interactions, which are not magnetically confined, can escape from the source (``neutron model'').
\item {\bf Direct escape regime.} Directly escaping protons from the outer edges of the shells dominate the UHECR injection, at least at the highest energies. 
\item {\bf Optically thick to neutron escape regime.} Only neutrons from the outer edges of the shells can escape.
\end{enumerate}
In case (1), one (muon) neutrino per cosmic ray will be produced; in case (2), the UHECR escape will not be necessarily accompanied by  neutrino production; and in case (3), the neutrino production will be enhanced compared to case (1), since the neutrinos can escape from everywhere within the shell.  Normally, case (1) has been assumed in the literature.

As one of our main results, we have identified the GRB parameter regions which are associated with these three different cases. We have demonstrated that
direct escape is, for instance, important for GRBs with large $\Gamma$, such as we have shown for several typical  {\it Fermi}-LAT observed bursts. The actual magnitude of the direct escape component somewhat depends on the acceleration efficiency, since protons can only escape directly at the highest energies, where the Larmor radius is of the order of the shell width. For efficient proton acceleration, we have demonstrated that the standard case (1), one neutrino per cosmic ray, only applies to a very narrow region of the parameter space at the highest energies, since either direct escape dominates, or the optical thickness to neutron escape is large. For less efficient proton acceleration, a significant region where the standard assumption (1) applies has been found, which is around the often assumed standard parameter values. However, we have illustrated that the maximal proton energies are in that case not sufficient to describe the observed UHECR spectrum for typical burst parameters.
 Therefore, it appears that the standard case (1) is in tension with the assumption that GRBs are the sources of the UHECRs, and specific conclusions can be only drawn on a burst-by-burst basis. Note, however, that even in the direct-escape dominated cases, the neutron escape will contribute at lower energies as well, which leads to a spectral break in the cosmic ray injection spectrum.

The region in which the optical thickness to neutron escape is large, case (3), has been found to be larger than previous calculations suggest, because high-energy processes have been included in the photohadronic interaction rate. In this case, the neutrinos can escape from everywhere within the shell, whereas the neutrons (and protons) are trapped over the photohadronic interaction length scale and can only (directly) escape from the edges. The neutrino production can therefore be significantly enhanced. 
Furthermore, we have explicitly demonstrated that the formula used for the pion production efficiency in the IceCube treatment in \citet{Abbasi:2009ig,Abbasi:2012zw} in fact underestimates the neutrino production in the optically thick case, and that the original formula in \citet{Guetta:2003wi} applies instead (which was originally developed for the optically thin case). This has consequences for individual GRBs, and is, in fact, already taken into account in the prediction by \citet{Hummer:2011ms}. 

From the model perspective, we have described the direct escape of protons and neutrons in a unified framework, and we have shown how the expansion of the fireball affects the escape. For instance, we have demonstrated that for a relativistic plasma with an adiabatic index  $\hat \gamma = 4/3$ the ratio between shell width and direct escape mean free path is independent of the shell radius, which means that the conclusions do not depend on the details of the time evolution. Note that there may be additional escape components, such as from diffusion, which however require a time-dependent calculation, whereas the direct escape component is guaranteed in a rather model-independent way.  Such additional components would however not affect our main arguments.

As far as the consequences for the UHECR observations are concerned, we have demonstrated that a two-component escape model, which includes cosmic rays from direct proton and neutron escape, can reproduce the regions around the dip and ankle extremely well, because the transition between the two components leads to a spectral break in the cosmic ray spectrum from a single source. Since an additional diffusive escape component may affect  this, and the effect depends on the GRB parameter set, it remains to be seen if such an extension of the dip model survives in a parameter space study.
 We expect that future neutrino observations will provide stringent constraints on two-component models using both the source (PeV) and cosmogenic (EeV) neutrino fluxes. While the source neutrino flux is correlated with the neutron production, the cosmogenic neutrino flux is insensitive to the escape mechanism of the cosmic rays; a detailed discussion of this issue will be performed elsewhere.

Finally, we note that the three different encountered regimes may even be present in one source, especially if collisions occur at very different radii. Therefore, the one neutrino per cosmic ray assumption is, in fact, not as general as one may believe. Note that some of our conclusions can be transferred to other classes of sources, such as active galactic nuclei, and to heavier nuclei accelerated in the sources. In all those cases, a substantial fraction of particles may directly leak from the sources at the highest energies, and there can be regions where the source is optically thick to baryon escape. 

\acknowledgments{

We are grateful to Amyad Spector and Eli Waxman, who consulted us in numerous enlightening discussions.
We would also like to thank Markus Ahlers, Francis Halzen, Svenja H{\"u}mmer, Kohta Murase, and Nathan Whitehorn for useful discussions regarding this work. PB and WW would also like to thank the Weizmann institute for their warm hospitality during a research visit, when this work was initiated.

WW would like to acknowledge support from DFG grant WI 2639/3-1.  MB and PB would like to acknowledge support from the GRK 1147 ``Theoretical Astrophysics and Particle Physics''. 
This work has also been supported by the FP7 Invisibles network (Marie Curie
Actions, PITN-GA-2011-289442), the ``Helmholtz Alliance for Astroparticle Physics HAP'', funded by the Initiative and Networking fund of the Helmholtz association, and DFG grant WI 2639/4-1.}

\appendix

\section{On the effects of diffusion}
\label{app:diffusion}

The description of diffusion depends on the actual magnetic field configuration, relevant scattering processes, and properties of the plasma. This means that a model-independent approach is not possible and a dedicated treatment of the transport equations, such as a set of Fokker-Planck equations, is required. However, we make some estimates for the additional effects of diffusion in this appendix, and demonstrate that they do not affect our qualitative conclusions.

The diffusion length $\lambda'_{\text{diff}}$,
over which the particles can escape, is given by
\begin{equation}
\lambda'_{\text{diff}} \simeq \sqrt{D \, t_{\mathrm{dyn}}'} \, ,
\label{equ:diff}
\end{equation}
on the dynamical timescale $t_{\mathrm{dyn}}'$.
Here $D$ is the diffusion coefficient $D \equiv L'^2/T'$, where $L'$ is the displacement scale and $T'$ time between the collisions relevant for diffusion. Since acceleration is assumed to be present over $t_{\mathrm{dyn}}'$, energy losses will take over thereafter, and conclusions on the additional escape after $t_{\mathrm{dyn}}'$ can only come from a time-dependent calculation. 

Now, the diffusion coefficient depends on the actual field configuration. For example, if the magnetic field is aligned on scales larger than the Larmor radius, photohadronic and other scattering processes can lead to diffusion perpendicular to the magnetic field. If the scattering by photohadronic interactions dominates and energy losses can be neglected, $D \simeq R_L'^2/t'_{p\gamma}$ and therefore
\begin{equation}
\lambda'_{\text{diff}} \simeq R_L' \sqrt{\frac{t_{\mathrm{dyn}}'}{t'_{p\gamma}}} \simeq  R_L' \, \sqrt{\tau_n} \, 
\label{equ:diffexample}
\end{equation}
at the highest energies.
This corresponds to the direct escape component, enhanced by the optical thickness $\sqrt{\tau_n}$ (\cf, \Sec~\ref{sec:thick}), since the photohadronic scattering processes lead to the transport of additional protons closer to the shell boundaries. From \figu{optthick}, however, one can easily see that this additional enhancement does not affect the qualitative conclusion that neutron escape dominates in the optically thick regime. 

Another regime which may be relevant for the diffusive escape in GRBs is the case of Bohm-like diffusion in magnetic fields. Assuming that our particles move with $c$, and re-writing the result in terms of $R'_L$, we obtain $D \approx R_L' c \left(\propto B'^{-1}\right)$.  In order to study the impact of such a component, we assume that \equ{mfp} is changed into
\begin{equation}
	\lambda'_{\text{mfp}}(E') =  \min \left[ \Delta r', \sqrt{R'_L(E') \, c \, t_{\mathrm{dyn}}'}, c\, t'_{p\gamma} (E') \right] \, , \label{equ:mf2}
\end{equation}
where \equ{diff} has been used to compute the fraction of diffusively escaping protons. This form implies that at most all of the protons can escape, and that this diffusive component will be suppressed if photohadronic interactions lead to significant energy losses or scattering. Note that other energy losses and time-dependent effects have been neglected here, which means that this can only serve as a rough estimate. 
In addition, it is implied that the amount of electrostatic turbulence in the plasma is only moderate. For a higher turbulence, actual MHD calculations are needed.
This means that corrections and even a factor-of-a-few deviations are quite likely, and even the energy dependence may change if that assumption does not hold. 

\begin{figure*}[pt]
  \centering
  \includegraphics[width=0.8\textwidth]{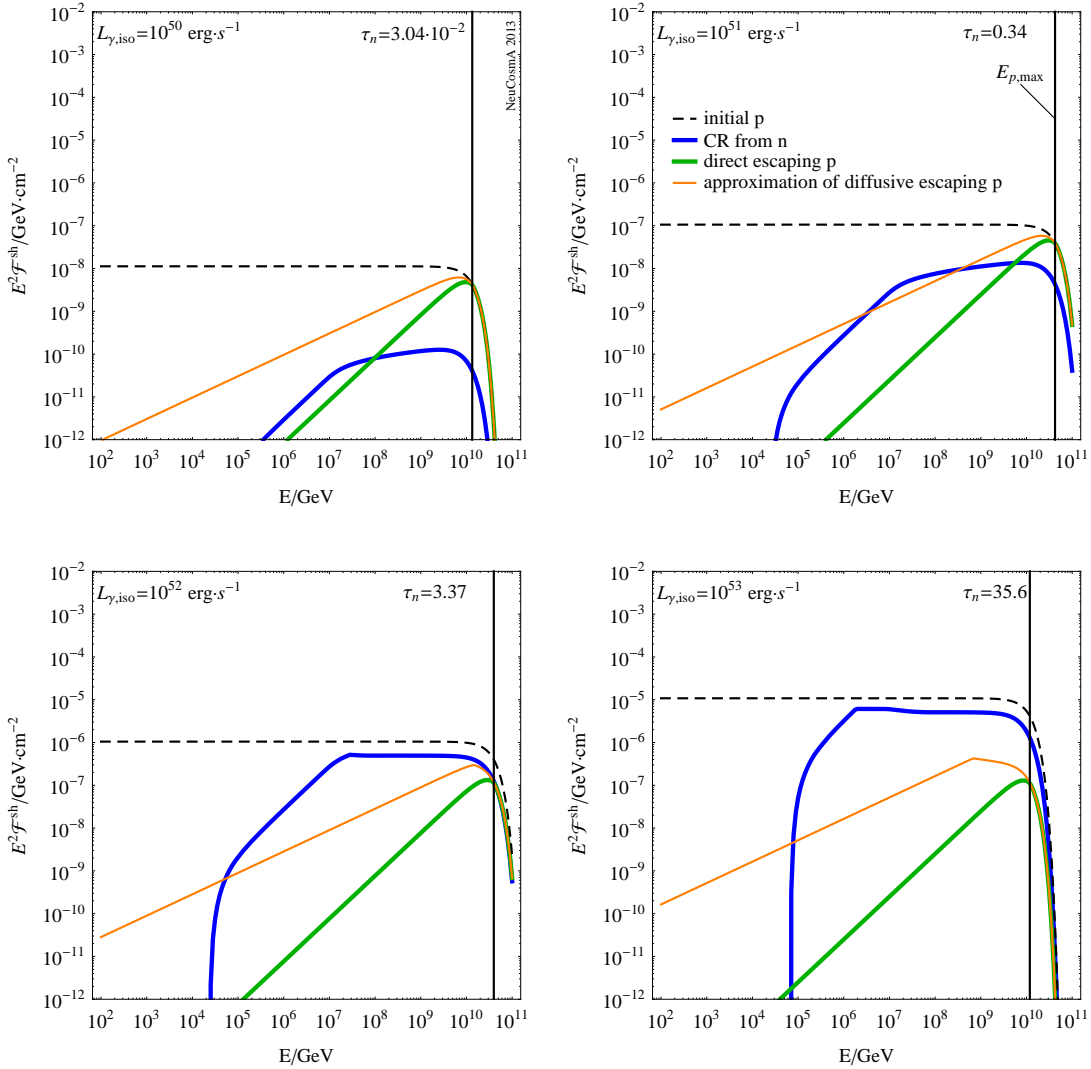}
  \caption{\label{fig:diffesc} Cosmic ray flux components as in \figu{optthin} and \figu{optthick}, including a (model-dependent) estimate for a diffusive escape component of protons (orange curves); see main text for details. 
}
\end{figure*}

We show in \figu{diffesc} the impact of the additional diffusive escape component, given our assumptions, for our standard examples. For the optically thin sources (to neutron escape) in the upper row, the diffusive component may be as large as the direct escape at the highest energies, \ie, almost all protons can escape. This is expected, since direct escape and diffusion describe the same limit there and the Larmor radius is comparable to the dynamical timescale at the highest energies. However, since $\lambda'_{\text{diff}}  \propto \sqrt{R_L'} \propto \sqrt{E'}$, the energy-dependence is different from direct escape, which means that at lower energies the diffusive escape may actually dominate.
For the optically thick sources in the lower row, the diffusive escape can be enhanced compared to the direct escape because $t_{\mathrm{dyn}}' > R_L'$ at the highest energies (\cf, \figu{optthick}, lower panels, where the acceleration rate, corresponding to $c \, R_L'^{-1}$, is larger than  $t_{\mathrm{dyn}}'^{-1}$). The spectral breaks come from the points where photohadronic interactions take over.
However, the neutron escape component still dominates above $10^5 \, \giga\electronvolt$.

This example illustrates that an additional diffusive escape component does not affect our conclusions, since (a) it does not change the results significantly at the highest energies and (b) it can only enhance the ejection of protons from the sources, which means that the region where the one neutrino per cosmic ray connection holds will be further reduced. However, it also demonstrates that our direct escape component can only be regarded as a lower estimate for the actual cosmic ray leakage from the sources.
As far as the impact on cosmic ray observations in \Sec~\ref{sec:cr} is concerned, the qualitative conclusions (additional component can better reproduce dip model) remain unchanged, but the actual energy dependence may depend on the interplay of different escape components. Since the estimates in this appendix are quite model-dependent, we do not mix them with the main text.

\section{Pion production efficiency in the optically thick regime}
\label{app:pion}

There have been several slightly different definitions of $f_\pi$ in the literature; see, \eg, \citep{Waxman:1997ti,Guetta:2003wi,Abbasi:2009ig}. In words, $f_\pi$  is defined as the fraction of proton energy going into pion production due to photohadronic interactions. For a single particle, this fraction has to be $f_\pi \leq 1$, whereas for a density of particles, this is no longer obvious and depends on the exact definition. One self-consistent possibility is to use the injection rates $Q'$ for the calculation, leading to the estimate
\begin{equation}
	f_\pi^{(1)} \equiv \frac{Q_\pi'}{Q_p'} \simeq \frac{t'^{-1}_{p\gamma} \cdot N_p' \cdot \langle x_\pi \rangle}{Q_p'} \quad .
	\label{equ:fpifromrates}
\end{equation}
Here, we use the fact that the pions are produced from the protons by photohadronic interactions at a rate $t'^{-1}_{p\gamma}$ with a pion inheriting the average amount of energy $\langle x_\pi \rangle \simeq 0.2$ from the parent proton.  In the optically thin case $t'^{-1}_{\text{dyn}} \geq t'^{-1}_{p\gamma}$, the particle injection counters the escape over the dynamical timescale in the steady state, and we get $Q_p' = t'^{-1}_{\text{dyn}} \cdot N_p'$ from \equ{steadystateesc}.
 When we insert this for $Q_p'$ into \equ{fpifromrates} we obtain 
\begin{equation}
	f_{\pi,\text{thin}}^{(1)} = \frac{t'^{-1}_{p\gamma}}{t'^{-1}_{\text{dyn}}} \, \langle x_\pi \rangle \leq \langle x_\pi \rangle < 1 \quad ,
	\label{equ:fpiQoptthin}
\end{equation}
which is exactly the result used by \citet{Guetta:2003wi}. 

However, this is no longer valid in the optically thick case $t'^{-1}_{\text{dyn}} < t'^{-1}_{p\gamma}$. As one of the subtleties, one may want to take into account multiple interactions. If the neutron decay rate can be neglected, as it is typically the case, protons and neutrons can be treated as one baryon system with similar interaction rates for all baryons. In that case, the particles are confined in this baryon system, and  photohadronic energy losses dominate. One can easily solve the kinetic equation describing continuous energy losses for a power law spectrum, in order to obtain $Q_p' = t'^{-1}_{p\gamma} \cdot N_p' \cdot \langle x_\pi \rangle $. By inserting this for $Q_p'$ in \equ{fpifromrates}, the result for $f_\pi$ simplifies to
\begin{equation}
	f_{\pi,\text{thick}}^{(1)} \simeq 1 \quad ,
	\label{eq:fpiQoptthick}
\end{equation}
\ie, the baryons lose all their energy into pion production.
Hence, when using the injection rates $Q'$ for the definition of $f_\pi$, no values larger than one can be obtained and conservation of energy is shown. The definition from Guetta et al.~\citep{Guetta:2003wi} at least correctly describes the optically thin case, where injection rates are used for the definition of $f_\pi$. 

On the other hand, a second possibility to define $f_\pi$ is implied  by the IceCube method to relate the neutrino and photon fluences in \citet{Abbasi:2009ig}, used for all current state-of-the-art stacking analyses. Here, the fluences are related to each other by energy partition arguments, which correspond to energy partition of the particle densities in the source. The pion production efficiency $f_\pi$ is calculated in the conventional way~\citep{Guetta:2003wi}, modified such that $f_\pi \le 1$ in the optically thick case. However, the definition of $f_\pi$ implied in this approach is based on  the particle (steady state) densities $N'$:
\begin{equation}
	f_\pi^{(2)} \equiv \frac{N_\pi'}{N_p'} \quad .
	\label{equ:fpifromdensities}
\end{equation}
For $N_\pi'$, we can estimate that $N_\pi' \approx Q_\pi' \cdot t'_{\text{dyn}}$ as long as the dynamic escape rate is larger than the decay rate, $t'^{-1}_{\text{dyn}} > t'^{-1}_{\text{dec}}$, \ie, for large enough energies.
For the sake of simplicity, we neglect synchrotron losses here, which of course leads to an additional correction of the pion production efficiency, as discussed earlier in \citet{Hummer:2011ms} as well as \citet{Li:2011ah} (included in $c_S$ in \citet{Hummer:2011ms}). This however does not change the argument here, but is included in our numerical calculations. In the optically thin case, we again have $N_p' \simeq Q_p' \cdot t_{\text{dyn}}'$, leading to
\begin{equation}
	f_{\pi,\text{thin}}^{(2)} \simeq \frac{Q_\pi' \cdot t_{\text{dyn}}'}{Q_p' \cdot t_{\text{dyn}}'} = \frac{Q_\pi'}{Q_p'} \quad ,
	\label{equ:fpiNoptthin}
\end{equation}
which is equivalent to the definition in \equ{fpifromrates} and leads to the result in \equ{fpiQoptthin}. It therefore does not make a difference what approach is used. However, in the optically thick case, we again have $N_p' \approx Q_p' \cdot t_{p\gamma}' \cdot \langle x_\pi \rangle ^{-1}$ (for the baryon system), and we obtain
\begin{equation}
	f_{\pi,\text{thick}}^{(2)} \simeq \frac{Q_\pi' \cdot t_{\text{dyn}}'}{Q_p' \cdot t_{p\gamma}'}  \langle x_\pi \rangle = \frac{Q_\pi'}{Q_p'} \cdot \frac{t'^{-1}_{p\gamma}}{t'^{-1}_{\text{dyn}}}  \langle x_\pi \rangle  = \frac{t'^{-1}_{p\gamma}}{t'^{-1}_{\text{dyn}}}  \langle x_\pi \rangle  \quad , 
	\label{equ:fpiNoptthick}
\end{equation}
where we have used in the last step that $Q'_\pi/Q'_p \simeq 1$ in the optically thick case from above. Clearly, since $t'^{-1}_{p\gamma} \gg t'^{-1}_{\text{dyn}}$ is possible, $f_\pi^{(2)}$ can be larger than one. It is interesting that the form in \equ{fpiNoptthick} is the same as in \citet{Guetta:2003wi}, which means that this formula applies in the optically thin and thick cases, if the definition in \equ{fpifromdensities} is used.\footnote{Note, however, that the original formula was not derived for that case: it was defined with respect to the definition in \equ{fpifromrates}, and did not include secondary interactions.}  Since this applies to the IceCube treatment, the formula for $f_\pi$ in \citet{Abbasi:2009ig} can be easily adjusted to describe the optically thick regime correctly.
The large neutrino flux in \figu{optthick} reflects the previously underestimated neutrino production in the optically thick case, whereas neutrons are trapped by photohadronic interactions. Note that numerical predictions, such as \citet{Hummer:2011ms}, include this enhancement automatically, but still find an order of magnitude lower predictions than with the original method. The reason is that not all bursts are affected by this. Especially two of the five most dominant bursts (of the original method) of the IC-40 sample~\citep{Abbasi:2011qc} are severely affected by this, which changes how these bursts contribute to the total flux. These two bursts have a neutrino flux which is enhanced by a factor of more than four in our approach. Another burst is mildly enhanced, \ie, by about 10\%, while the other two originally dominating bursts are unaffected by this. As a result, only three of the original five bursts dominate in the new analysis.

Now one may ask: where does this qualitative difference come from, and why does that not violate energy conservation? If the energy partition arguments are applied to the actual densities in the source (and not the injection spectra), as it may be the more plausible approach at least for electrons and photons, the secondary production can become very large since it is proportional to the product of proton and photon densities; \cf, \equ{prodmaster}. The proton density in the source will be balanced by the injection $Q_p' = t'^{-1}_{p\gamma} \cdot N_p' \cdot \langle x_\pi \rangle $ in the steady state in the optically thick case.
That is, for fixed $N_p'$ (from energy partition), it is implied that more protons have to be  injected if $t'^{-1}_{p\gamma}$ increases. Thus, the definition in \equ{fpifromdensities} is with respect to an open box to which energy will be injected from the outside, whereas the definition in \equ{fpifromrates} includes energy conservation automatically. Since in the optically thick case the maximal proton energy is typically limited by the photohadronic interaction rate, the proton acceleration will always be more efficient than the energy losses up to the maximal proton energy. 
This situation can be expected in the collision phase. If, however, the acceleration is switched off at some point, the proton density will decay quickly and energy conservation applies again. Note that by using the actual proton and photon densities as input for the photohadronic interactions, we automatically have the optically thick case included if we define energy equipartition with respect to the baryon system (\ie, do not distinguish between protons and neutrons in the source). In practice, there will be about 50\% of protons and 50\% of neutrons in a source with a large optical thickness, according to the secondary multiplicities of protons and neutrons. 
For neutrinos from optically thick sources, only observables sensitive to the $\pi^+/\pi^-$ ratio will be sensitive to the actual distribution between protons and neutrons, such as the Glashow resonance; see discussion in \citet{Hummer:2010ai}. 


\end{document}